%% file: ZL.tex
\documentclass{aa}
\usepackage{epsfig}

\begin{document}

\title{Optical spectrum of the planetary nebula M\,2-24}
\author{Y. Zhang\inst{1}\thanks{\emph{Present address:}
        Department of Astronomy, Peking University,
	        Beijing 100871, P. R. China}
		 \and X.-W. Liu\inst{2}}

\offprints{Y. Zhang,\\
 \email{zhangy@bac.pku.edu.cn}}

 \institute{National Astronomical Observatories, Chinese Academy of Sciences,
            Beijing 100012, P. R. China
     \and Department of Astronomy, Peking university, Beijing 100871,
          P. R. China}

\date{Received/Accepted}


\abstract{

We have obtained medium-resolution, deep optical long-slit spectra of the
bulge planetary nebula (PN) \object{M\,2-24}. The spectrum covers the
wavelength range from 3610--7330~{\AA}. Over two hundred emission lines have
been detected.  The spectra show a variety of optical recombination lines
(ORLs) from C, N, O and Ne ions. The diagnostic diagram shows significant
density and temperature variations across the nebula. Our analysis
suggests that the nebula has a dense central emission core. The nebula was thus
studied by dividing it into two regions: 1) an high ionization region
characterized by an electron temperature of $T_{\rm e}=16\,300$\,K and a
density of $\log\,N_{\rm e}({\rm cm}^{-3}) = 6.3$; and 2) a low
ionization region represented by $T_{\rm e}=11\,400$\,K and $\log\,N_{\rm
e}({\rm cm}^{-3}) = 3.7$. A large number of ORLs from C, N, O and Ne
ions have been used to determine the abundances of these elements relative to
hydrogen. In general, the resultant abundances are found to be higher than
the corresponding values deduced from collisionally excited lines (CELs).
This bulge PN is found to have large enhancements in two
$\alpha$-elements, magnesium and neon.

}

\maketitle

\section{Introduction}

\object{M\,2-24} (\object{PK\,356--5\degr\,2}) is a compact (angular radius
$\sim4$\,arcsec), faint [$\log F({\rm
H}\beta)=-12.10$\,(erg\,cm$^{-2}$\,s$^{-1}$); Cahn et al.  \cite{cahn1992}]
planetary nebula in the Galactic bulge. Beaulieu et al. (\cite{beaulieu})
measured its heliocentric radial velocity as 137\,km\,s$^{-1}$.  Using a
statistical method, Zhang (\cite{zhang}) obtained a distance of 11.79\,kpc for
\object{M\,2-24}. Similar to \object{IC\,4997} (Hyung et al. \cite{hyung1994}),
the spectrum of \object{M\,2-24} shows an extremely large [\ion{O}{iii}]
$\lambda4363$ to H$\gamma$ $\lambda4340$ intensity ratio, suggesting that it
may have a high-density emission core. The presence of large density
inhomogeneities in a nebula could lead to erroneous elemental abundances
determined using the empirical method based on collisionally excited lines
(CELs; Zhang \& Liu \cite{zhangliu}). 

The determination of abundances in PNe provide very important constraints
on the nuclear and mixing processes in their progenitor stars and on the
chemical evolution of galaxies.  Some recent chemical abundance studies of PNe
in the Galactic Bulge are given by Ratag et al. \cite{ratag92}, Ratag et al.
\cite{ratag97}, Cuisinier et al. \cite{cuisinier}, Escuder \& Costa
\cite{escuder}. However, in all these studies, the heavy element abundances
relative to hydrogen were based on CELs only.

We have obtained medium-resolution, deep optical long-slit spectra of the
bulge PN \object{M\,2-24}. A large number of optical recombination lines
(ORLs) detected in its spectrum can be used to determine elemental abundances.
Unlike CELs, ionic abundances derived from ORLs are insensitive to the electron
temperature and density in the emission region. Thus the presence of a
high-density emission core in \object{M\,2-24} will hardly affect the
abundances determined from ORLs. A long-standing problem in nebular abundance
studies has been that the heavy-element abundance derived from ORLs are
systematically higher than those derived from CELs (see Liu
\cite{liua,liub} for recent reviews). As we show in this paper, even in
the PNe with very dense cores, the large density fluctuation cannot
completely explain the discrepancies between the ORL and CEL abundances.  Based
on detailed studies of several PNe, Liu (\cite{liua}) concluded that the large
discrepancies between the ORL and CEL abundances observed in some nebulae
are mostly caused by the presence of extremely cold H-deficient inclusions
embedded in normal nebular material.

In Sect.~2 of this paper, we describe our new optical observations 
of \object{M\,2-24} and data
reduction and present the observed line fluxes. Dust extinction towards 
\object{M\,2-24} is discussed in Sect.~3. Plasma diagnostic analyses
are presented in Sect.~4. In Sect.~5, we present and discuss the
ionic and elemental abundances derived from both ORLs and CELs.
A general discussion then follows 
in Sect.~6.


\section{Observations and data reduction}

\begin{table}
\caption{Observational journal}
\label{jan}
\centering
\begin{tabular}{ccccc}
\hline
\hline
\noalign{\smallskip}
Date &$\lambda$-range &Slit Width& FWHM &Exp. Time\\
     & ({\AA}) &(arcsec)& ({\AA}) & (sec)\\
\noalign{\smallskip}
\hline
\noalign{\smallskip}
1996 Jul&3530--7430&2&4.5&60, 300\\
1996 Jul&3530--7430&8&4.5&60\\
2001 Jun&3500--4805&2&1.5&900, 1800\\
\noalign{\smallskip}
\hline
\end{tabular}
\end{table}

The observations were carried out with the ESO 1.52\,m telescope using the
long-slit spectrograph Boller \& Chivens (B\&C).  A journal of observations is
presented in Table~\ref{jan}.  In 1996, the detector was a UV-enhanced Loral
$2048\times 2048$ $15\,\mu{\rm m}\times 15\,\mu{\rm m}$ chip, which was
superseded in 2001 by a $2688\times 2688$ $15\,\mu{\rm m}\times 15\,\mu{\rm m}$
chip. The B\&C spectrograph has a useful slit length of about 3.5\,arcmin. In
order to reduce the  CCD read-out noise, in 1996 the CCD was binned by a factor
of two along the slit direction, yielding a spatial sampling of
1.63~arcsec per pixel projected on the sky. A slit width of 2\,arcsec was used
throughout except for one short exposure for which an 8\,arcsec wide slit was
used so that total line fluxes from the whole nebula were recorded.  The slit
was oriented in the north-south direction (PA$=0^\circ$). The wavelength range
from 3530\,{\AA}--7430\,{\AA} was  observed in 1996 with an effective
spectral resolution of 4.5~{\AA}, as determined from the FWHM of the
calibration lamp lines. A short exposure was taken in order to obtain
intensities of the brightest emission lines, which were saturated in the
spectrum of 5\,min exposure time.  At the same slit position an
additional wavelength range was observed in 2001, covering
3500\,{\AA}--4805\,{\AA} at a resolution of 1.5~{\AA} FWHM. 

\begin{figure*}
\vspace{0.4cm}
 \centering \epsfig{file=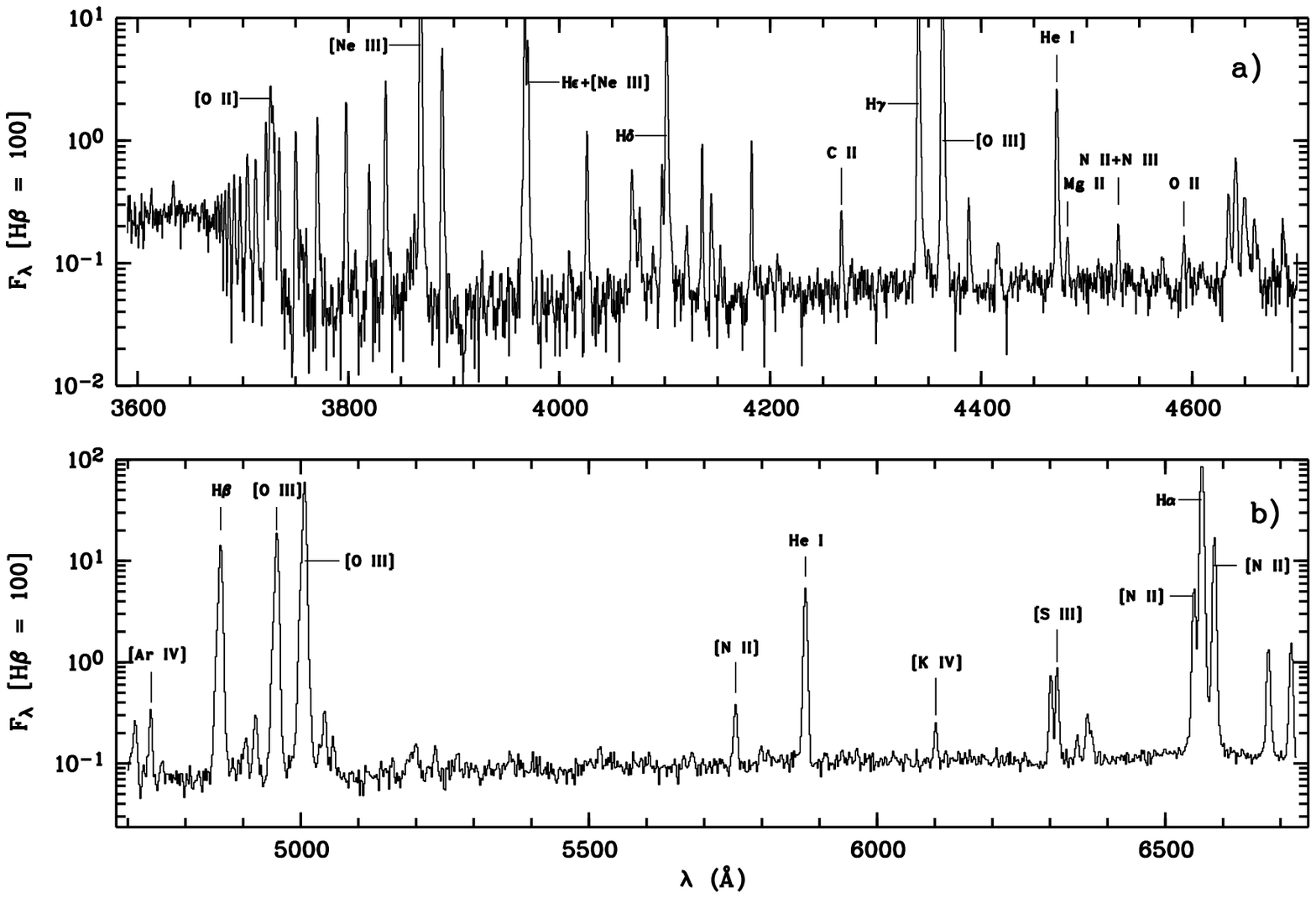,
height=11.5cm, bbllx=39, bblly=256, bburx=546, bbury=600, clip=, angle=0}
\caption{Optical spectrum of M\,2-24 from a) 3600--4700
{\AA} and b) 4700--6700 {\AA}.
Note that H$\alpha$ and [\ion{O}{iii}] $\lambda$5007
were saturated in this exposure. Interstellar extinction has not been
corrected for.}
\label{spectrum}
\end{figure*}

All the spectra were reduced using the {\sc long92} package in {\sc
midas}\footnote{{\sc midas} is developed and distributed by the European
Southern Observatory.} following the standard procedure. The spectra were
bias-subtracted, flat-fielded and cosmic-rays removed, and then wavelength
calibrated using exposures of a HeAr-CuFe calibration lamps.  Absolute flux
calibration was obtained by observing the standard stars \object{Feige\,110}
and the nucleus of \object{PN NGC\,7293} (Walsh \cite{walsh}). The extracted
spectra \object{M\,2-24}, after integrating along the slit, are plotted
in Fig.~\ref{spectrum}. The spectra have not been corrected for interstellar
extinction. The spectrum taken with an 8\,arcsec wide slit yields an
${\rm H}\beta$ flux of $\log F({\rm H}\beta) = -12.11$
(erg\,cm$^{-2}$\,s$^{-1}$), which is in good agreement with the value of -12.10
(erg\,cm$^{-2}$\,s$^{-1}$) tabulated in Cahn et al. (\cite{cahn1992}).

All line fluxes, except those of the strongest lines, were measured using Gaussian
line profile fitting. For the strongest and isolated lines, fluxes were obtained
integrating over the observed line profiles.
A full list of observed lines and their measured fluxes are presented in
Table~\ref{linelist}. In Table~\ref{linelist}, column (1) gives the observed
wavelengths after corrected for the Doppler shift as determined from the Balmer
lines. The observed fluxes are given in column (2). Column (3) lists the
fluxes after corrected for interstellar extinction, 
$I(\lambda)=10^{cf(\lambda)}F(\lambda)$, where $f(\lambda)$ is
the standard Galactic extinction law for a total-to-selective
extinction ratio of $R=3.1$ (Howarth \cite{howarth}) and $c$ is the
logarithmic extinction at ${\rm H}\beta$ (cf. Sect.~3).
Column (4)--(10) give the ionic identification,
laboratory wavelength, multiplet number, the lower and upper spectral terms of
the transition, and the statistical weights of the lower and upper levels,
respectively. All fluxes are normalized such that ${\rm H}\beta = 100$.

\input{tab1.tex}

\section{Reddening summary}
\begin{table}
\caption{\label{extin}Extinction derived from the Balmer decrement}
\centering
\begin{tabular}{ccc}
\hline
\hline
\noalign{\smallskip}
Balmer decrement & Ratio & $c({\rm H}\beta)$\\
\noalign{\smallskip}
\hline
\noalign{\smallskip}
 H$\alpha$/H$\beta$&6.16&1.08\\
H$\gamma$/H$\beta$&0.39&0.65\\
H$\delta$/H$\beta$&0.20&0.70\\
Adopted&&0.80\\
\noalign{\smallskip}
\hline
\end{tabular}
\end{table}

The measured H$\alpha$/H$\beta$, H$\gamma$/H$\beta$, and
H$\delta$/H$\beta$ ratios, together with the logarithmic extinction at
H$\beta$, $c=\log I({\rm H}\beta)/F({\rm H}\beta)$, derived from these ratios
using the Galactic reddening law of Howarth (\cite{howarth}), are listed in
Table~\ref{extin}.  The extinction derived from the H$\alpha$/H$\beta$ ratio is
higher than those derived from the higher order Balmer lines, suggesting
possible self-absorption effects caused by significant optical depths in the
Balmer lines. The Balmer line ratios H$\alpha$/H$\beta$ and H$\gamma$/H$\beta$
as a function of the optical depths of Ly\,$\alpha$ and H\,$\alpha$ are given
by Cox \& Mathews (\cite{cox}). For an H\,$\alpha$ optical depth of 5, the
observed H$\alpha$/H$\beta$ and H$\gamma$/H$\beta$ ratios yield very similar
reddening constants of $c({\rm H}\beta) = 0.80$.  For comparison, the observed
total H$\beta$ flux, $\log F({\rm H}\beta)=-12.11$\,erg\,cm$^{-2}$\,s$^{-1}$
and the 5-GHz radio free-free continuum flux density, $S(5\,{\rm
GHz})=0.003$\,Jy, for \object{M\,2-24} (Cahn et al. \cite{cahn1992}) yield a
very low value, $c({\rm H}\beta) = 0.08$. It is possible that the radio flux
density for this faint PN may have been significantly underestimated, leading
to an underestimated interstellar extinction constant.  On the other hand, it
is also possible, as has been known for some time (Stas\'{i}ska et al.
\cite{stasinska}; Walton, Barlow \& Clegg \cite{walton}; Liu et al.
\cite{liu2001}), the standard interstellar extinction law ($R_V = 3.1$) may not
be applicable to the lines-of-sight towards Galactic bulge PNe.  Here we
have adopted $c({\rm H}\beta) = 0.80$, as derived from the observed Balmer line
ratios after corrected for the optical depth effects, to deredden the optical
spectra.

\section{Plasma diagnostics}

\begin{figure*}
\centering
\epsfig{file=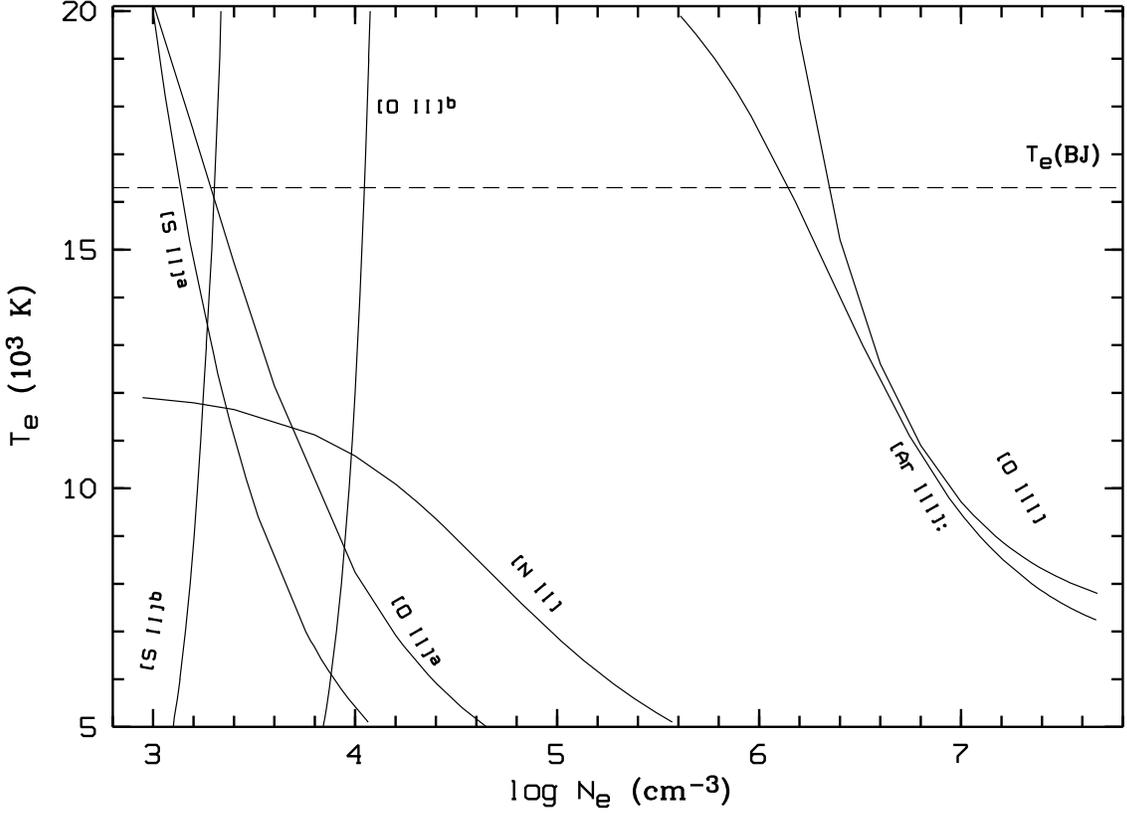, height=11.0cm,
bbllx=68, bblly=419, bburx=514, bbury=744, clip=, angle=0}
\caption{Plasma diagnostic diagram for M\,2-24. The dashed line shows the
hydrogen Balmer jump temperature.}
\label{tnplane}
\end{figure*}

In the optical spectrum of \object{M\,2-24}, a number of CELs, which are useful
for nebular diagnostic analysis and abundance determinations, are observed.
The electron temperatures and densities derived from various CEL diagnostic
ratios were obtained by solving the level populations for multilevel ($\geq 5$)
atomic models. The plasma diagnostic diagram based on these line ratios are
plotted in Fig.~\ref{tnplane}. The ratios employed were [\ion{N}{ii}]
$(\lambda6548+\lambda6584)/\lambda5754$, [\ion{O}{ii}]$^a$
$(\lambda7320+\lambda7330)/(\lambda3726+\lambda3729)$, [\ion{O}{ii}]$^b$
$\lambda3726/\lambda3729$, [\ion{S}{ii}]$^a$
$(\lambda4068+\lambda4076)/(\lambda6731+\lambda6716)$, [\ion{S}{ii}]$^b$
$\lambda6731/\lambda6716$, [\ion{O}{iii}]
$(\lambda4959+\lambda5007)/\lambda4363$ and [\ion{Ar}{iii}]
$\lambda7135/\lambda5192$. Note that the intensity of the weak line
[\ion{Ar}{iii}] $\lambda5192$ has a relatively large uncertainty making the
[\ion{Ar}{iii}] $\lambda7135/\lambda5192$ ratio less reliable as a plasma
diagnostic.   According to the intersectant points of the temperature
diagnostic [\ion{N}{ii}] $(\lambda6548+\lambda6584)/\lambda5754$ with four
density diagnostics [\ion{O}{ii}]$^a$
$(\lambda7320+\lambda7330)/(\lambda3726+\lambda3729)$, [\ion{O}{ii}]$^b$
$\lambda3726/\lambda3729$, [\ion{S}{ii}]$^a$
$(\lambda4068+\lambda4076)/(\lambda6731+\lambda6716)$ and [\ion{S}{ii}]$^b$
$\lambda6731/\lambda6716$, we can conclude that the singly ionized regions can
be characterized by a temperature of $T_{\rm e} = 11\,400\pm500$\,K and a
density of $\log N_{\rm e}({\rm cm}^{-3}) = 3.7\pm0.4$. On the other hand,
emission from doubly ionized species such as [\ion{O}{iii}]
$(\lambda4959+\lambda5007)/\lambda4363$ appear to arise from a separate high
density emission region.

\begin{table*}
\caption{Plasma diagnostics}
\label{diagnostic}
\centering
\begin{tabular}{llccc}
\hline
\hline
\noalign{\smallskip}
Ion& Lines & Ionization Potential\,(eV)& Ratio &Result\\
\noalign{\smallskip}
\hline
\noalign{\smallskip}
 & & & &$\log N_{\rm e}$\,(cm$^{-3}$)\\
{[\ion{S}{ii}]}& $I(\lambda6731)/I(\lambda6716)$  & 10.36 & 1.32 & 3.25$^{\mathrm{a}}$\\
{[\ion{S}{ii}]}& $I(\lambda4068+\lambda4076)/I(\lambda6731+\lambda6716)$& 10.36 & 0.22& 3.38$^{\mathrm{a}}$\\              
{[\ion{O}{ii}]}& $I(\lambda3726)/I(\lambda3729)$  &13.62 &2.36 & 3.99$^{\mathrm{a}}$\\
{[\ion{O}{ii}]}& $I(\lambda7320+\lambda7330)/I(\lambda3726+\lambda3729)$ &13.62 &0.13 & 3.67$^{\mathrm{a}}$\\ 
{[\ion{Ar}{iii}]}& $I(\lambda7135)/I(\lambda5192^{\mathrm{b}})$ & 27.63&33.2& 6.15$^{\mathrm{c}}$\\
{[\ion{O}{iii}]} & $I(\lambda4959+\lambda5007)/I(\lambda4363)$ & 35.12 &6.56& 6.35$^{\mathrm{c}}$\\
\multicolumn{2}{l}{Balmer decrement}   &     &       & 6--7\\
\noalign{\vskip5pt}
 & & & &$T_{\rm e}$\,(K)\\
{[\ion{N}{ii}]}& $I(\lambda6548+\lambda6584)/I(\lambda5754)$ & 14.53 & 59.55 & 11400$^{\mathrm{d}}$\\
 BJ/H\,11 &     &         &       & 16300 \\
\noalign{\smallskip}
\hline
 \end{tabular}
 \begin{list}{}{}
 \item[$^{\mathrm{a}}$] Assuming $T_{\rm e}=11\,400$\,K;
 \item[$^{\mathrm{b}}$] Poor quality;
 \item[$^{\mathrm{c}}$] Assuming $T_{\rm e}=16\,300$\,K;
 \item[$^{\mathrm{d}}$] Assuming $\log N_{\rm e}({\rm cm}^{-3})=3.7$.
 \end{list}
 \end{table*}

\begin{figure}
\centering
\epsfig{file=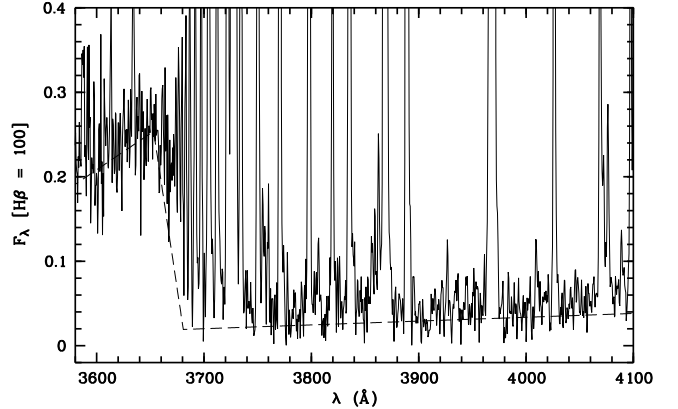, height=5.5cm,
bbllx=102, bblly=426, bburx=468, bbury=657, clip=, angle=0}
\caption{Spectrum of M\,2-24 from 3580--4100\,{\AA},
showing the nebular continuum Balmer discontinuity at 3646\,{\AA}.
The dashed line is an empirical fit to the continuum level. 
The spectrum has not been corrected for extinction and is normalized such 
that
$F({\rm H\beta})=100$.}
\label{bal}
\end{figure}

Fig.~\ref{tnplane} also shows the Balmer jump (BJ) temperature
derived from the ratio of the nebular
continuum Balmer discontinuity at 3646\,{\AA} to H\,11 $\lambda3770$ 
(Fig.~\ref{bal}), using the formula given by Liu et al. (\cite{liu2001}).
We used the ratio of the Balmer discontinuity to H~11 rather than 
to H$\beta$, since the temperature thus derived is much less sensitive to 
uncertainties in the reddening correction and flux calibration, given the 
small wavelength difference between the Balmer discontinuities and H~11.
The He$^+$/H$^+$ and He$^{++}$/H$^+$ ionic abundance ratios, which are
needed to calculate $T_{\rm e}$(BJ), are taken from Sect.~5. 
We found $T_{\rm e}(\rm{BJ})=16\,300$\,K, which is about 5000\,K higher 
than the temperature implied by the nebular diagnostics from singly 
ionized species.

If we assume that the [\ion{O}{iii}] and [\ion{Ar}{iii}] lines arise from
a hotter region of $T_{\rm e}(\rm{BJ})=16\,300$\,K, then the
[\ion{O}{iii}] and [\ion{Ar}{iii}] diagnostic ratios plotted in 
Fig.~\ref{tnplane} would yield an electron density of 
$\log N_{\rm e}({\rm cm}^{-3})=6.35$ and 6.15, respectively. This would suggest
that this hotter, high ionization region (dominated by doubly ionized species)
is probably also much denser than the cooler region from which lines from
singly ionized species arise.

\begin{figure}
\centering
\epsfig{file=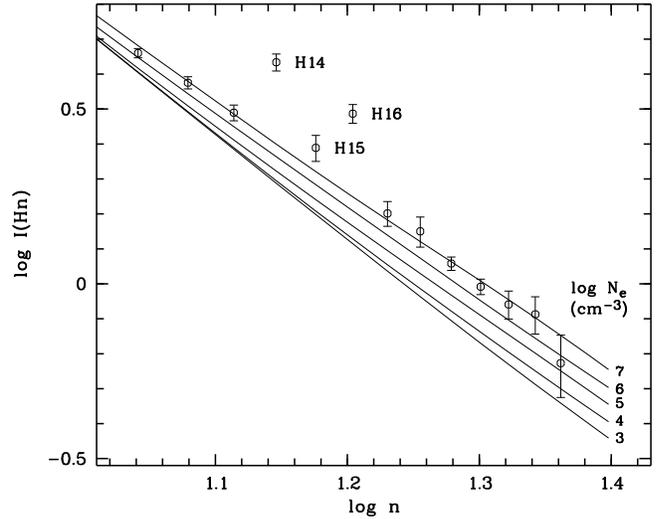, height=7.0cm,
bbllx=43, bblly=419, bburx=444, bbury=742, clip=, angle=0}
\caption{Observed intensities (in units where ${\rm H}\beta=100$) of 
the high-order Balmer lines ($n\rightarrow2$, $n=11,13,...,23$) as a
function of the principal quantum number $n$. 
Note H~14, H~15 and H~16 are affected by line blending. The solid lines
show predicted intensities for electron densities
from $N_{\rm e}=10^3$ to 10$^7$\,cm$^{-3}$, assuming an electron temperature
of 16\,300\,K.}
\label{re_ne}
\end{figure}

That \object{M\,2-24} contains a dense emission region is supported by our
analysis of its hydrogen recombination line spectrum.  In
Fig.~\ref{re_ne}, we plot the observed intensities of high-order hydrogen
Balmer lines ($n\rightarrow2$, $n=11,13,...,23$) as a function of the principal
quantum number $n$ of the upper level.  Theoretical intensities for different
electron densities are also plotted assuming an electron temperature of
16\,300\,K, as deduced from the ratio of nebular continuum Balmer discontinuity
to H~11. The intensities of the Balmer decrement are density-sensitive but
depend only weakly on  temperature, thus provide a valuable density diagnostic.
In particular, they can be used to probe high-density ionization regions where
collisionally excited diagnostic lines of relatively low critical densities are
suppressed by collisional de-excitation. Given our spectral resolution,
the Balmer lines can be resolved (and thus the Balmer decrement can be
measured) up to $n=23$.  Fig.~\ref{re_ne} shows that the measured intensities
of almost all the Balmer lines from $n=11$ to 23 yield an best fit density
between $10^6$--$10^7$\,cm$^{-3}$,  except for H~14 $\lambda$3721.94, H~15
$\lambda$3711.97 and H~16 $\lambda$3703.86, which are blended with the
[\ion{S}{iii}] $\lambda3721.63$, \ion{O}{iii} $\lambda3715.08$ and \ion{He}{i}
$\lambda3705.12$ lines, respectively.

\begin{figure}
\centering
\epsfig{file=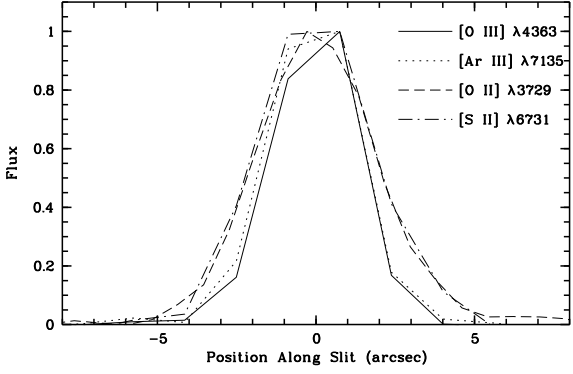, height=5.0cm,
bbllx=101, bblly=427, bburx=457, bbury=658, clip=, angle=0}
\caption{Spatial distribution of emission along the slit for [\ion{O}{iii}]
$\lambda4363$, [\ion{Ar}{iii}] $\lambda7135$, [\ion{O}{ii}]
$\lambda3729$ and [\ion{S}{ii}] $\lambda6731$ lines. For all the emission
lines, the peak-values of the observed fluxes are normalized to one.
}
\label{profile}
\end{figure}

Table~\ref{diagnostic} summarizes these electron temperatures and
densities derived from the various diagnostics, which suggests that
[\ion{O}{iii}] and [\ion{Ar}{iii}] emission lines arise probably from a
different region with those singly ionized species. In Fig.~\ref{profile}, we
compare the surface brightness distributions of several emission lines along
the slit.  The figure shows that the spatial distribution of [\ion{O}{iii}]
$\lambda4363$ line is similar to that of the [\ion{Ar}{iii}] $\lambda7135$
line, while emissions from singly ionic species such as the [\ion{O}{ii}]
$\lambda3729$ and [\ion{S}{ii}] $\lambda6731$ lines are apparently from more
extensive regions.

The above analysis reveals that \object{M\,2-24} has two distinct emission
regions of very different physical conditions. The hotter and denser region,
characterized by $T_{\rm e}=16\,300$\,K and $\log N_{\rm e}({\rm cm}^{-3})=6.3$,
is from a dense core close to the central star.
The cooler, lower density region, characterized 
by  $T_{\rm e}=11\,400$\,K and $\log N_{\rm e}({\rm cm}^{-3})=3.7$ and 
dominated by emission from singly ionized species, is from an outer
shell. Thus \object{M\,2-24} bears many resemblances to \object{Mz\,3}
previously analyzed by us (Zhang \& Liu \cite{zhangliu}).

\begin{figure}
\centering
\epsfig{file=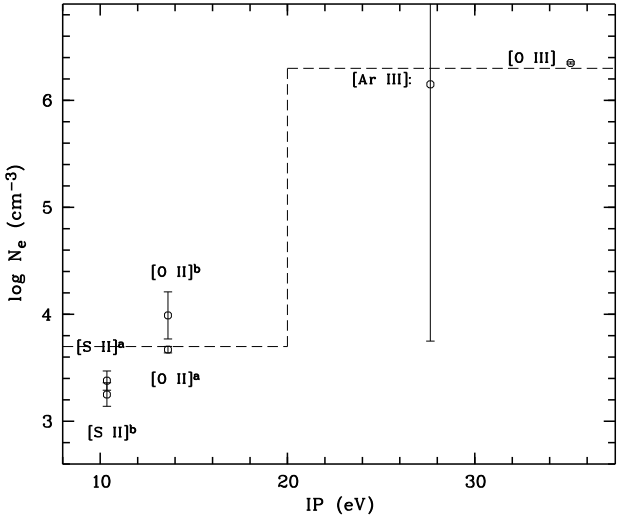, height=7.0cm,
bbllx=62, bblly=419, bburx=443, bbury=743, clip=, angle=0}
\caption{Electron density versus ionization potential. The dashed line shows 
the electron densities adopted to determine the relative ionic abundances.}
\label{ipne}
\end{figure}

Fig.~\ref{ipne} illustrates
$N_{\rm e}$ derived from various diagnostics as a function of 
Ionization Potential (IP) of the diagnostic lines. 
For the purpose of ionic abundance determinations (see below), we have thus
divided the nebula into two
regions based on IP, as indicated by the dashed line in Fig.~\ref{ipne}.

\section{Abundance analysis}
In this section we present the elemental abundances
derived from observed line intensities relative to H$\beta$. 
For both CELs and ORLs abundance calculations, the effective recombination coefficient 
of H$\beta$ was taken from Storey \& Hummer (\cite{storey1995}).
Based on the plasma diagnostic results, a constant temperature of
$T_{\rm e}=16\,300$\,K and a
density of $\log N_{\rm e}({\rm cm}^{-3})=6.3$ have been assumed for
ions with IP higher than 20\,eV. For
ions of lower IP, a constant temperature of 
$T_{\rm e}=11\,400$\,K
and a density of $\log N_{\rm e}({\rm cm}^{-3})=3.7$ have been assumed.

In Sect.~3, we showed that the Balmer line emission from the nebula
is probably optically thick and estimated an optical depth 
of 5 for H$\alpha$. The effect of self-absorption on H$\beta$ flux has to be
corrected for before the observed line intensities tabulated in 
Table~\ref{linelist} can be used to calculate the ionic abundances.
The escape
probability of H$\beta$ as a function of H$\alpha$ optical depth has been
calculated by Cox \& Mathews (\cite{cox}). By interpolating their results, we derived
an escape probability of 0.67 for H$\beta$ for an H$\alpha$ optical depth of 5.
Consequently, we have divided all the line intensities relative to H$\beta$
tabulated in Table~\ref{linelist}  by a factor of 0.67 before calculating the
ionic abundances.

\subsection{Ionic abundances from CELs}

The ionic abundances are derived by \begin{equation} \frac{N({\rm
X}^{i+})}{N({\rm H}^+)}=\frac{I_{jk}}{I_{{\rm
H}\beta}}\frac{\lambda_{jk}}{\lambda_{{\rm H}\beta}}\frac{\alpha_{{\rm
H}\beta}}{A_{jk}}({\frac{N_j}{N({\rm X}^{i+})}})^{-1}N_{\rm e} \end{equation}
where $I_{jk}/I_{{\rm H}\beta}$ is the ratio of the intensity of the ionic line
to H$\beta$; $\lambda_{jk}/\lambda_{{\rm H}\beta}$ is the ratio of the
wavelength of the ionic line to that of H$\beta$; $\alpha_{{\rm H}\beta}$ is
the effective recombination coefficient of ${\rm H}\beta$; $A_{jk}$ is the
Einstein spontaneous transition probability of the ionic line; and $N_j/N({\rm
X}^{i+}$) is the ratio of the fractional population of the upper level from
which the ionic line originates, which is sensitive to the electron temperature
and density. References for the adopted collision strengths and transition
probabilities are listed in Table~\ref{abund_re}.

\begin{table}
\caption{\label{abund_re}Reference for atomic data of CELs.}
\centering
\begin{tabular}{lccll}
\hline
\hline
\noalign{\smallskip}
Ion & \multicolumn{2}{c}{IP\,(eV)} & \multicolumn{2}{c}{Reference}\\
X$^i$&  X$^{i-1}$ & X$^i$  & Trans. Prob. & Coll. Str.  \\
\noalign{\smallskip}
\hline
\noalign{\smallskip}
N$^+$ & 14.53 & 29.60& [1] & [2] \\
O$^+$ & 13.62&35.12&[3] & [4] \\
O$^{2+}$&35.12&54.94 & [5] & [6] \\
Ne$^{2+}$&40.90&63.45 & [7] & [8] \\
S$^+$ &10.36&23.34&  [9], [10] & [11] \\
S$^{2+}$ &23.34&34.79& [12] & [7] \\
Ar$^{2+}$&27.63&40.74 & [13] & [14] \\
Ar$^{3+}$&40.74&59.81 & [9] & [15] \\
Fe$^{2+}$&16.19&30.65 & [16]  & [17] \\
Fe$^{3+}$&30.65&54.80 & [18]      & [19] \\
\noalign{\smallskip}
\hline
\end{tabular}
\begin{list}{}{}
 \item[References:] [1] Nussbaumer \& Rusca (\cite{nussbaumer}); [2] Stafford et al. 
(\cite{stafford}); [3] Zeippen (\cite{zeippen1982}); [4]Pradhan (\cite{pradhan});
[5] Nussbaumer \& Storey (\cite{nussbaumer1981}); [6] Aggarwal (\cite{aggarwal}); 
[7] Mendoza (\cite{mendoza1983}); [8] Butler \& Zeippen (\cite{butler1994});
[9] Mendoza \& Zeippen (\cite{mendoza1982a}); [10] Keenan et al. (\cite{keenan1993});
[11] Keenan et al. (\cite{keenan1996}); [12] Mendoza \& Zeippen (\cite{mendoza1982b}); 
 [13] Mendoza \& Zeippen (\cite{mendozazeippen1983}); [14] Johnson \& Kingston 
(\cite{johnson1990}); [15] Zeippen et al. (\cite{zeippen1987}); [16]
Nahar \& Pradhan(\cite{nahar}); [17] Zhang (\cite{zhang1996}); [18]
Zhang \& Pradhan (\cite{zhang1997}); [19] Garstang (\cite{garstang1958}).
\end{list}
\end{table}

The ionic abundances derived from optical CELs are presented in
Table~\ref{cel_ab}.  Given that ionic abundances derived from CELs are very
sensitive to the adopted electron temperature and density, and the fact that
their values in the dense central emission core of \object{M\,2-24} are not
well constrained by our current observations, the results presented in
Table~\ref{cel_ab} may suffer large uncertainties.  Additional errors may also
be introduced by the fact that not all the observed fluxes of lines with IP $>$
20~eV arise entirely within the dense core. Conversely, not all the observed
fluxes of lines with IP $<$ 20~eV arise exclusively in the the outer, less
dense nebular layers. In other words, our simple two-zone model may not be
sufficient for such a complicated object.

For comparison, Table~\ref{cel_ab}
also gives the N$^+$/H$^+$, O$^+$/H$^+$ and S$^+$/H$^+$ abundance ratios derived
from the [\ion{N}{ii}] $\lambda$5754, [\ion{O}{ii}] $\lambda\lambda$7320, 7330
and [\ion{S}{ii}] $\lambda\lambda$4068, 4076 auroral or transauroral lines, 
which, owning to their higher critical densities, are less density-dependent 
than the [\ion{N}{ii}] $\lambda\lambda$6548, 6584, 
[\ion{O}{ii}] $\lambda\lambda$3726, 3729
and [\ion{S}{ii}] $\lambda\lambda$6716, 6731 nebular lines. On the other hand,
given their higher excitation energies, the ionic abundances derived from
auroral or transauroral lines are more sensitive to the adopted $T_{\rm e}$ than those
derived from nebular lines. The abundances derived from
the auroral or transauroral lines are in good agreement with those deduced from the
nebular lines, suggesting the adopted electron temperature and density for the
low ionization region are reasonable.

\begin{table}
\caption{\label{cel_ab} Ionic abundances of heavy elements derived from CELs.}
\centering
\begin{tabular}{llr}
\hline
\hline
\noalign{\smallskip}
Ion& Lines & $N_{{\rm X}^{i+}}/N_{{\rm H}^+}$\\
\noalign{\smallskip}
\hline
\noalign{\smallskip}
N$^+$       & [\ion{N}{ii}] $\lambda\lambda$6548, 6584 & 5.25(-6)\\
            & [\ion{N}{ii}] $\lambda\lambda$5754 & 4.86(-6)\\
O$^+$       & [\ion{O}{ii}] $\lambda\lambda$3726, 3729 & 3.61(-6)\\
            & [\ion{O}{ii}] $\lambda\lambda$7320, 7330 & 4.82(-6)\\
O$^{2+}$    &[\ion{O}{iii}] $\lambda\lambda$4959, 5007&8.04(-5)\\
Ne$^{2+}$   &[\ion{Ne}{iii}] $\lambda3868$&2.24(-5)\\
S$^+$       &  [\ion{S}{ii}]  $\lambda\lambda$6716, 6731& 2.57(-7)\\
            &  [\ion{S}{ii}]  $\lambda\lambda$4068, 4076& 1.63(-7)\\
S$^{2+}$    &  [\ion{S}{iii}] $\lambda6312$& 4.08(-7)\\
Ar$^{2+}$   & [\ion{Ar}{iii}] $\lambda7135$ &1.93(-7)\\
Ar$^{3+}$   & [\ion{Ar}{iv}] $\lambda4733$ &2.46(-7)\\
Fe$^{2+}$   &  [\ion{Fe}{iii}] $\lambda\lambda$4658, 4734 & 8.71(-8)\\
Fe$^{3+}$   &  [\ion{Fe}{iv}] $\lambda$5032 & 1.10(-6)\\
\noalign{\smallskip}
\hline
\end{tabular}
\end{table}

\subsection{Ionic abundances from ORLs}
In order to compare with the ionic abundances derived from CELs, we 
also determine ionic abundances using ORLs.
From the measured intensities of ORLs, ionic
abundances can be derived using
\begin{equation}
\frac{N({\rm X}^{i+})}{N({\rm H}^+)}=\frac{I_{jk}}{I_{{\rm H}\beta}}\frac{\lambda_{jk}}{\lambda_{{\rm H}\beta}}\frac{\alpha_{{\rm H}\beta}}{\alpha_{jk}}
\end{equation}
where $\alpha_{jk}$ is the effective recombination coefficient for the ionic
ORL. Ionic abundances derived from ORLs depend only weakly on the adopted temperature,
$\sim T_{\rm e}^\alpha$, where $|\alpha|\la1$, and are
essentially independent of $N_{\rm e}$ under the low density conditions
(N$_e\la10^8$~cm$^{-3}$). 
As a consequence, the presence of temperature and density fluctuations will
have little effects on ORL abundances.
The ionic abundances derived from ORLs are presented in Table~\ref{orl_ab}.
ORLs blended with strong lines have large uncertainties in the measured 
intensities and therefore have been excluded from the table.

\begin{table}
\caption{\label{orl_ab} Ionic abundances of M\,2-24 from ORLs.}
\centering
\begin{tabular}{lllcc}
\hline
\hline
\noalign{\smallskip}
Ion &$\lambda$\,(\AA)  & Mult &  $I$           & $N_{{\rm X}^{i+}}/N_{{\rm H}^+}$\\
\noalign{\smallskip}
\hline
\noalign{\smallskip}
He$^+$  & 4471.50    & V14  & 6.885          & 6.74(-2) \\
        & 5875.66    & V11  & 20.66          & 6.41(-2) \\
        & 6678.16    & V46  & 4.470          & 7.00(-2) \\
        & Average    &      &                & 6.59(-2) \\
He$^{2+}$& 4685.68   &  3.4 & 0.559          & 3.46(-4) \\
\noalign{\vskip8pt}
C$^{2+}$& 4267.15    & V6  & 0.657          & 4.42(-4)\\
C$^{3+}$&  4186.90     & V18   & 0.162        & 1.29(-4)\\
\noalign{\vskip8pt}
N$^{2+}$&  5679.56     & V3    & 0.217        & 2.56(-4)\\
        &4607.16$^{\mathrm{a}}$     & V5    & 0.080        & 7.32(-4)\\
        &5940.24,1.65& V28   & 0.142        & 3.49(-4)\\
        &4035.08     & V39a  & 0.089        & 3.16(-4)\\
        &4041.31     & V39b  & 0.065        & 1.95(-4)\\
        &4236.91,7.91& V48   & 0.060        & 1.77(-4)\\
        &4432.74,3.48& V55   & 0.034        & 1.84(-4)\\
        &4678.14     & V61b  & 0.054        & 2.66(-4)\\
        &Average     &       &              & 3.00(-4)\\
N$^{3+}$&4379.11$^{\mathrm{b}}$  & V60b    & 0.121 & 3.73(-5)\\
\noalign{\vskip8pt}
O$^{2+}$& 4676.24    & V1   & 0.175          & 9.42(-4) \\
    &4661.63    & V1   & 0.102          & 4.61(-4) \\
    &4349.43    & V2   & 0.277          & 9.13(-4) \\
    &4317.14$^{\mathrm{c}}$   & V2   & 0.091          & 6.80(-4) \\   
    &4414.90$^{\mathrm{d}}$   & V5   & 0.133          & 2.11(-3) \\ 
    &4416.97$^{\mathrm{d}}$   & V5   & 0.220          & 3.48(-3) \\
    &4078.84    & V10  & 0.071          & 1.28(-3) \\
    &4072.16$^{\mathrm{e}}$   & V10  & 0.283          & 0.73(-3) \\   
    &3882.19,3.13$^{\mathrm{f}}$&V12  & 0.131          & 2.34(-3) \\  
    &4156.53$^{\mathrm{d}}$ & V19  & 0.102          & 5.52(-3) \\
    &4153.30    & V19  & 0.228          & 1.32(-3) \\
    &4089.29    & V48a & 0.275          & 1.96(-3)\\
    &4062.94$^{\mathrm{g}}$ & V50a & 0.052          & 2.91(-3)\\   
    &4048.21$^{\mathrm{g}}$ & V50b & 0.054          & 5.48(-3)\\   
    &4273-78    & V67  & 0.038          & 6.64(-4) \\
    &4281-84    & V53,67& 0.124         & 2.94(-3)\\
    &4288.82$^{\mathrm{d}}$& V53c & 0.081         & 5.70(-3)\\   
    &4332.71$^{\mathrm{d}}$ & V65b  & 0.117         & 3.55(-3)\\ 
    &4285.69    & V78b  & 0.041         & 1.49(-3)\\
    &4609.44,10.20& V92   & 0.127         & 1.58(-3)\\
    &Average    &         &               & 1.40(-3)\\
\noalign{\vskip8pt}
Ne$^{2+}$& 3777.14     & V1    & 0.197        & 6.82(-4)\\
       &4428.64,52  & V60c,V61c& 0.087     & 1.35(-3)\\
       &Average     &          &           & 1.02(-3)\\
\noalign{\vskip8pt}
Mg$^{2+}$&  4481.21     & V4    & 0.345        & 2.41(-4)\\
\noalign{\smallskip}
\hline
\end{tabular}
\begin{list}{}{}
  \item[$^{\mathrm{a}}$] Possibly blended with [\ion{Fe}{iii}] $\lambda4607.13$;
  \item[$^{\mathrm{b}}$] Corrected for 30\,per cent contributions 
  from the \ion{Ne}{ii} (V60b) $\lambda4379.55$ line,
  assuming \ion{Ne}{ii} $I(4379.55)/I(4391.99)=0.61$;
  \item[$^{\mathrm{c}}$] Includes a 8.6\,per cent contribution from 
\ion{O}{ii} $\lambda4317.70$;
  \item[$^{\mathrm{d}}$] Possibly contaminated by other lines; Discarded
 in the calculation of the average ionic abundances;
  \item[$^{\mathrm{e}}$] Includes a 6.2\,per cent contribution from
\ion{O}{ii} $\lambda4071.23$;
  \item[$^{\mathrm{f}}$] Includes a 16\,per cent contribution from
\ion{O}{ii} $\lambda 3882.45$;
\item[$^{\mathrm{g}}$] Poor quality; Discarded
 in the calculation of the average ionic abundances.
\end{list}
\end{table}

The He$^+$/H$^+$ abundance ratios derived from the $\lambda\lambda$4471, 5876
and 6678 lines were weighted by 1:3:1, roughly proportional to the intrinsic
intensity ratios of the three lines. In a typical PN, the singlet
resonance lines of \ion{He}{i} are strongly optically thick, thus Case B
recombination is usually a much better approximation than Case A for the
\ion{He}{i} singlet recombination lines.  The \ion{He}{i} triplets have no
$n=1$ ground level and resonance lines from the meta-stable (pseudo) ground
level 2s\,$^3$S are normally optically thin, thus Case A recombination should
be a good approximation for the \ion{He}{i} triplet lines. Therefore,
Case A recombination was assumed for the triplet lines $\lambda$4471 and
$\lambda$5876, and Case B for the singlet $\lambda$6678 line. A recent
calculation of the effective recombination coefficients of \ion{He}{i} lines is
given by Smits (\cite{smits1996}). The contributions to the observed line
fluxes by collisional excitation from the He$^0$ 2s$^3$S and 2s$^1$S
meta-stable levels by electron impacts were studied by Sawey \& Berrington
(\cite{sawey}).  Combining the recombination data of Smits (\cite{smits1996})
and the collision strengths of Sawey \& Berrington (\cite{sawey}), Benjamin et
al.  (\cite{benjamin}) presented improved values for \ion{He}{i} line emission
coefficients and fitted the results with analytical formulae. We have adopted
their formulae in determining the He$^+$/H$^+$ abundance ratios. The results
derived from the three lines, $\lambda\lambda$4471, 4876 and 6678, agree
reasonably well.  The He$^{2+}$/H$^+$ abundance ratio was calculated using the
\ion{He}{ii} $\lambda$4686 line only, for which the effective recombination
coefficient was taken from Storey \& Hummer (\cite{storey1995}). Compared with
He$^+$/H$^+$, the He$^{2+}$/H$^+$ abundance ratio is negligible, suggesting
\object{M\,2-24} is a relatively low excitation nebula.

In Table~\ref{heicom}, we compare the intensities of \ion{He}{i} lines
relative to $\lambda$4471 line observed in the spectrum of \object{M\,2-24} with the predictions
of Benjamin et al. (\cite{benjamin}) for an electron
temperature of $T_{\rm e}=11\,400$\,K and a density of
$\log\,N_{\rm e}({\rm cm}^{-3}) = 3.7$. The relative differences, 
$\Delta=(I_{\rm obs}-I_{\rm pred})/I_{\rm obs}$ are also given. 
Table~\ref{heicom} shows excellent agreement between the observations
and recombination theory for the 2p\,$^3$P$^{\rm o}$--$n$d\,$^3$D and
2p\,$^1$P$^{\rm o}$--$n$d\,$^1$D series. The observed intensity of
2s\,$^3$S--$3$p\,$^3$P$^{\rm o}$ $\lambda3889$ is much lower than
the theoretical prediction, suggesting absorption and reprocessing of photons
of the 2s\,$^3$S--$n$p\,$^3$P$^{\rm o}$ series to be quite significant. As a result, 
the 2p\,$^3$P$^{\rm o}$--$3$s\,$^3$S $\lambda7065$ is enhanced, consistent with
the observations. A detail discussion of the effect
of optical depth of the 2\,$^3$S level on the nebular spectrum of 
\ion{He}{i} line is given by Benjamin et al. (\cite{benjamin2002}).
Table~\ref{heicom} also shows that the observed intensities of the
2s\,$^1$S--$5$p\,$^1$P$^{\rm o}$ $\lambda3614$ and 
2p\,$^1$P$^{\rm o}$--$3$p\,$^1$S$^{\rm o}$ $\lambda7281$ are significantly lower
than the predicted values. Similar features have also been found in
the spectra of other PNe, such as \object{NGC\,6153} 
(Liu et al. \cite{liubarlow}), \object{M\,1-42}, and \object{M\,2-36}
(Liu et al. \cite{liu2001}). This is attributed 
to the destruction of
\ion{He}{i} Lyman photons by the photoionization of H$^0$ or by
absorption of dust grains (Liu et al. \cite{liu2001}).

\begin{table}
\caption{\label{heicom} Intensities of the \ion{He}{i} lines in
\object{M\,2-24}, normalized such that $I(4471)=1.00$, where
$\Delta=(I_{\rm obs}-I_{\rm pred})/I_{\rm obs}$. Case A has
been assumed for the triplets and Case B for the singlets. The 
results are compared to the theoretical values deduced from
Benjamin et al. (\cite{benjamin})}
\centering
\begin{tabular}{lcccr}
\hline
\hline
\noalign{\smallskip}
$\lambda_0$ & n & $I_{\rm obs}$ &$I_{\rm pred}$& $\Delta$ \\
\noalign{\smallskip}
\hline
\noalign{\smallskip}
\multicolumn{5}{c}{2s\,$^1$S--$n$p\,$^1$P$^{\rm o}$ }\\
3613.64 & 5 & 0.064 &0.103 & $-60.9$\%\\
\noalign{\vskip5pt}
\multicolumn{5}{c}{2p\,$^1$P$^{\rm o}$--$n$s\,$^1$S }\\
4437.55 & 5      &  0.019 & 0.016& $+15.8$\%\\
7281.35 & 3      &  0.097 & 0.200& $-106$\% \\
\noalign{\vskip5pt}
\multicolumn{5}{c}{2p\,$^1$P$^{\rm o}$--$n$d\,$^1$D }\\
4387.93 & 5      &  0.113 & 0.111& $+1.8$\%\\
4921.93 & 4      &  0.233 & 0.252& $-8.2$\%\\
6678.16 & 3      &  0.649 & 0.741& $-14.2$\%\\
\noalign{\vskip5pt}
\multicolumn{5}{c}{2s\,$^3$S--$n$p\,$^3$P$^{\rm o}$ }\\
3888.65$^{\mathrm{a}}$& 3 & 0.659  & 2.598& $-294$\% \\
\noalign{\vskip5pt}
\multicolumn{5}{c}{2p\,$^3$P$^{\rm o}$--$n$s\,$^3$S }\\
7065.25 & 3      &  1.317 & 0.979&$+25.7$\% \\
\noalign{\vskip5pt}
\multicolumn{5}{c}{2p\,$^3$P$^{\rm o}$--$n$d\,$^3$D }\\
4026.21$^{\mathrm{b}}$ & 5 & 0.425 & 0.422& $+0.7$\%\\
4471.50 & 4      &  1.000 & 1.000 &   0.0\%\\
5875.66 & 3      &  3.001 & 2.917 &   $+2.8$\%\\
\noalign{\smallskip}
\hline
\end{tabular}
\begin{list}{}{}
  \item[$^{\mathrm{a}}$]Corrected for a 69\,per cent contributions
from H~8 $\lambda3889.05$ line using $I({\rm H}~8)/I({\rm H}~7)=0.66$;
  \item[$^{\mathrm{b}}$] Corrected for a 1.8\,per cent contributions 
from the \ion{O}{ii} (V 39b) $\lambda4026.08$ line using 
\ion{O}{ii} $I(\lambda4026.08)/I(\lambda4041.31)=0.82$.
\end{list}
\end{table}

The C$^{2+}$/H$^+$ abundance ratio was estimated from the 
3d\,$^2$D--4f\,$^2$F $\lambda$4267 doublet, which has a high S/N ratio. 
The adopted effective recombination
coefficient for the \ion{C}{ii} transition was from Davey et al. (\cite{davey}),
considering the effects of both radiative and di-electronic recombination.
Case B was assumed, although
the effective recombination coefficient of the \ion{C}{ii} $\lambda$4267 line is 
fairly case-insensitive.
The C$^{3+}$/H$^+$ abundance ratio was derived from the
$\lambda4187$ (V~18) \ion{C}{iii} recombination line based on the 
effective recombination coefficient given by P\'{e}quignot et al. 
(\cite{pequignot}) and the di-electronic recombination coefficient given
by Nussbaumer \& Storey (\cite{nussbaumer1984}). The \ion{C}{iv} 3s\,$^2$S--3p\,$^2$P$^{\rm o}$
$\lambda\lambda5802, 5812$ lines of multiplet V~1 have been detected in the
spectrum of \object{M\,2-24}.
However, no effective recombination coefficient is available for
this multiplet. Thus ORL C$^{4+}$/H$^+$ abundance ratio cannot be
determined. In fact, the \ion{C}{iv} $\lambda\lambda5802, 5812$ lines
are expected to arise mainly in the wind from the central star due to their
broad feature and the low He$^{2+}$/H$^+$ abundance ratio in the PN.

The N$^{2+}$/H$^+$ abundance ratio was derived from \ion{N}{ii} ORLs from the
3s--3p, 3p--3d and 3d--4f configurations. Only triplet lines were detected and
Case B recombination was assumed for all these triplets.  The strongest
transition is V3 3s\,$^3$P$^{\rm o}$--3p\,$^3$D $\lambda5679.56$ and its
effective recombination coefficient is fairly insensitive to the assumption of
Case A or B. In comparison, the effective recombination coefficient of the
other strong multiplet, V\,28 3p\,$^3$P--3d\,$^3$D$^{\rm o}$ $\lambda5941$, is
extremely case-sensitive. Comparison of N$^{2+}$/H$^+$ abundances derived from
the two multiplets, assuming Case B, suggests Case B is a good
approximation for V\,28. The latest calculation for the effective
recombination coefficients of \ion{N}{ii} lines, including contributions from
both radiative and di-electronic recombination, is given by Kisielius \& Storey
(\cite{kisielius2002}), assuming LS-coupling. The calculation however did not
include the 3d--4f transitions for which effects due to departure from
LS-coupling become important. We have thus adopted the effective recombination
coefficients given by Kisielius \& Storey (\cite{kisielius2002}) for the 3s--3p
and 3p--3d transitions and those by Escalante \& Victor (\cite{escalante}) for
the 3d--4f transitions.  In the spectra of \object{M\,2-24} and
\object{M\,2-36}, Liu et al. (\cite{liu2001}) found that the N$^{2+}$/H$^+$
ratios derived from the 3s--3p (V~3 and V~5) transitions are significantly
higher than those derived from the 3d--4f lines and attributed the discrepancy
to continuum fluorescence excitation of the 3--3 transitions by starlight. The
fluorescence can play a role in the excitation of the 3--3 transitions, but not
for transitions from the 3d--4f configurations. In \object{M\,2-24}, the ionic
abundance derived from the \ion{N}{ii} multiplet V3 3s\,$^3$P$^{\rm
o}$--3p\,$^3$D $\lambda5680$ is in good agreement with those deduced from the
3d--4f configurations, suggesting that fluorescence is not important in this
nebula. The \ion{N}{ii} V~5 3s\,$^3$P$^{\rm o}$--3p\,$^3$P $\lambda4607$ line
yields a higher N$^{2+}$/H$^+$ ratio than the average (by a factor of 2--3),
but this is probably caused by its blending with the [\ion{Fe}{iii}]
$\lambda4607$ line. The N$^{3+}$/H$^+$ abundance ratio was derived from the
\ion{N}{iii} $\lambda4379$ line of Multiplet V18 using the effective
recombination coefficient of P\'{e}quignot et al.  (\cite{pequignot}) and the
di-electronic recombination coefficient of Nussbaumer \& Storey
(\cite{nussbaumer1984}).  Although a number of \ion{N}{iii} lines of Multiplet
V3 are detected, no effective recombination coefficients are available for
these lines. \ion{N}{iii} lines of Multiplets, V1 and V2 are excited by Bowen
fluorescence mechanism or by stellar continuum fluorescence, thus cannot
be used for abundance determinations.

A number of \ion{O}{ii} multiplets have been detected, both doublets and 
quartets. The O$^{2+}$/H$^+$ abundance ratios derived from them 
agree reasonably well except for a few cases. The effective recombination 
coefficients are from
Storey (\cite{storey1994}) for 3s--3p transitions (assuming 
$LS$-coupling) and
Liu et al. (\cite{liu1995}) for 3p--3d and 3d--4f transitions (assuming intermediate
coupling). Case A was assumed for the doublets and Case B for the quartets.
However, all the multiplets analyzed here are case-insensitive except for V19. The good
agreement between the O$^{2+}$/H$^+$ abundance ratios derived from 
V19 and those from the other
multiplets suggests that Case B is a good assumption for V19. 
Several multiplets of \ion{O}{iii} permitted lines have been detected. Unfortunately, all of them
are dominated by excitation of the Bowen fluorescence mechanism or by the radiative
charge transfer reaction of O$^{3+}$ and H$^0$ (Liu \& Danziger 
\cite{liu1993}; Liu et al. \cite{liub1993}) rather than by recombination. 
Thus they cannot be used for abundance determinations.

Table~\ref{orl_ab} also gives the Ne$^{2+}$/H$^+$ abundance ratios 
derived from several \ion{Ne}{ii} ORLs. For the \ion{Ne}{ii} 
3s\,$^4$P--3p\,$^4$P $\lambda3777$ line (V~1), the effective recombination
coefficients was taken from Kisielius et al. (\cite{kisielius1998}), calculated
in $LS$-coupling. Case B was assumed although the 
multiplet is case-insensitive. For the 3d--4f lines, 
preliminary effective recombination coefficients calculated in 
intermediate coupling (Storey, private communication) were 
used. As in the case of \object{NGC\,6153} (Liu et al. \cite{liubarlow}), 
\object{NGC\,7009}(Luo et al. \cite{luo}), \object{M\,1-42} and \object{M\,2-36} 
(Liu et al. \cite{liu2001}), the Ne$^{2+}$/H$^+$ ratio derived from the \ion{Ne}{ii}
3s\,$^4$P--3p\,$^4$P $\lambda3777$ multiplet is lower than those deduced from the
3d-4f transitions. For \object{M\,2-24}, the ratio of the average Ne$^{2+}$/H$^+$ ratio
derived from the 3d--4f lines to the value deduced from the 3s--3p line
is 2.0, comparable with the corresponding values of 1.7, 2.0, 1.4 and 2.0 for 
\object{NGC\,6153}, \object{NGC\,7009}, \object{M\,1-42} and \object{M\,2-36},
respectively. 
The cause of the disparity remains unknown.

Mg$^{2+}$/H$^+$ ratio has been derived from the \ion{Mg}{ii} 3d\,$^2$D--4f\,$^2$F$^{\rm o}$
$\lambda4481$ line. As pointed out by Barlow et al. (\cite{barlow}), 
the \ion{Mg}{ii} 4481.21 line 
is the strongest and easiest to measure ORL from
any third-row ion. The effective recombination coefficient of the
\ion{C}{ii} $\lambda4267$ has been assumed for the 
\ion{Mg}{ii} $\lambda4481$ line in the calculation of Mg$^{2+}$/H$^+$, 
given the similarity between the atomic structure of \ion{Mg}{ii} and \ion{C}{ii}. Mg$^{2+}$ 
occupies an unusually large
ionization potential interval, from 15\,eV to 80\,eV,
thus in a typical nebula, essentially all Mg exists in the form of Mg$^{2+}$. 

\subsection{Comparison of ORL and CEL abundances of O$^{2+}$ and Ne$^{2+}$}
O$^{2+}$/H$^+$ and Ne$^{2+}$/H$^+$ ratios are available from both ORLs and
CELs. In both cases, the ORL abundances are found to be significantly higher
than the corresponding  values derived from CELs. Peimbert et al.
(\cite{peimbert1993}) calculated O$^{2+}$/H$^+$ abundances ratios in two
\ion{H}{ii} regions the \object{Orion Nebula}, \object{M\,17} and in the
planetary nebula NGC\,6572 using \ion{O}{ii} ORLs and found that they are about
a factor of 2 higher than those derived from [\ion{O}{iii}] forbidden lines.
They attributed the discrepancy to the presence of spatial temperature
fluctuations. However, detailed abundance analyses for the PNe
\object{NGC\,7009} (Liu et al.  \cite{liu1995}; Luo et al. \cite{luo}),
\object{NGC\,6153} (Liu et al., \cite{liubarlow}), \object{M\,1-42} and
\object{M\,2-36}( Liu et al., \cite{liu2001}), which show extremely large
discrepancies between the ORL and CEL abundances, ranging from a factor of 5 up
to a factor of 20, have found that infrared fine-structure lines, which are
temperature-insensitive, also yield quite low ionic abundance similar to those
derived from the optical and UV CELs. Thus fluctuations of temperature are not
the main cause of the discrepancy. In \object{M\,2-24}, the discrepancies are a
factor of 17 and 46 for O$^{2+}$/H$^+$ and Ne$^{2+}$/H$^+$, respectively. 
Considering that \object{M\,2-24} shows an unusual large density contrast
(approx. 3 orders of magnitude) between the core and the low ionization
regions, the density fluctuations might have played an important role in
causing the discrepancy. In the dense region of \object{M\,2-24}, CELs are
strongly suppressed due to heavy collisional de-excitation.  The electron
density of $\log N{\rm e}({\rm cm}^{-3})=6.3$ adopted above may have been
underestimated for the [\ion{O}{iii}] and [\ion{Ne}{iii}] emission regions.  As
a result, the ionic abundances derived from these lines may have been
underestimated. The Balmer decrement suggests the electron density of
\object{M\,2-24} can be as high as $\sim10^7$\,cm$^{-3}$ (see
Fig.~\ref{re_ne}).  If we assume a density of $10^{7.3}$\,cm$^{-3}$, then the
O$^{2+}$/H$^+$ abundance ratio derived from CELs becomes consistent with that
from ORLs.  However, even for such a high density, the Ne$^{2+}$/H$^+$ ratio
derived from CELs increases by only a factor of about five, which is far below
a factor of 46, a factor of required to reconcile the ORL and CEL abundances.
Although [\ion{Ne}{iii}] lines might arise from regions of even higher
densities because of their higher critical densities than [\ion{O}{iii}] lines,
the required density turns out to be $10^{8.7}$\,cm$^{-3}$ in order to
reconcile the ORL and CEL Ne$^{2+}$/H$^+$ ratios, which is relatively higher
than the value suggested by Balmer decrement.  Therefore, density fluctuations
alone cannot account for the discrepancy, at least in case of
Ne$^{2+}$/H$^+$ abundances derived from lines of different excitation
mechanisms.

\subsection{Total abundance}

\begin{table*}
\caption{Elemental abundances in M\,2-24 from CELs and ORLs, in units
such that ${\rm log}\,N(H)=12.0$}
\label{abco}
\centering
\begin{tabular}{ccccccc}
\hline
\hline
\noalign{\smallskip}
Element& ICF(ORLs)& ICF(CELs) &ORLs & CELs   & Average$^{\mathrm{a}}$ & Solar$^{\mathrm{b}}$\\
\noalign{\smallskip}
\hline
\noalign{\smallskip}
He & 1.63& --  &11.03&   --   & 11.06   & 10.93\\
C  & 1.04& --  &8.77 &   --   & 8.74    &  8.52\\
N  & 1.05& 23.3&8.55 & 8.09   & 8.35    &  7.92\\
O  & 1.03& 1.00&9.16 & 7.92   & 8.68    &  8.83\\
Ne & 1.04& 1.04&9.03 & 8.37   & 8.09    &  8.08\\
Mg & 1.00& --  &8.38 &   --   &  --     &  7.58\\
S  & --  & 2.01& --  & 6.13   & 6.92    &  7.33\\
Ar & --  & 1.04& --  & 5.66   & 6.39    &  6.40\\
Fe & --  & 1.00& --  & 6.08   &  --     &  7.50\\
\noalign{\smallskip}
\hline
\end{tabular}
\begin{list}{}{}
 \item[$^{\mathrm{a}}$] The average abundances of Galactic PNe (Kingsburgh \& Barlow, \cite{kingsburgh});
 \item[$^{\mathrm{b}}$] Grevesse \& Sauval (\cite{grevesse});
\end{list}
\end{table*}

The total elemental abundances derived for \object{M\,2-24} from
CELs and ORLs are presented in Table~\ref{abco}. The ionization
correction factors (ICFs) listed in the second column of this table
are for the corrections of He, C, N, O, Ne and Mg ORL abundances and 
N, O, Ne, S, Ar and Fe CEL abundances (see below).
For comparison, we also list the average abundances deduced for Galactic PNe
by Kingsburgh \& Barlow (\cite{kingsburgh}) and
the solar photospheric abundances compiled by Grevesse \& Sauval 
(\cite{grevesse}).

Due to the relatively low excitation nature of \object{M\,2-24} (the
excitation class ${\rm E.C.}=1.7$, see Sect.~6), a significant amount
of neutral helium in the ionized hydrogen zone is expected. In order to
correct for the unseen He$^0$, it is necessary to evaluate the helium  ICF. For
this purpose, we note that He$^0$ has an ionization potential of 24.5\,eV, very
close to the value of 23.3\,eV for S$^+$. Thus to a good approximation, we have
\begin{eqnarray} \frac{{\rm He}}{{\rm H}}&=&{{\rm ICF}({\rm
He})}\times{\frac{{\rm He}^+}{{\rm H}^+}}={\frac{{\rm S}^++{\rm S}^{2+}}{{\rm
S}^{2+}}}\times{\frac{{\rm He}^+}{{\rm H}^+}}.\nonumber\\ \nonumber
\end{eqnarray} From the S$^+$ and S$^{2+}$ abundances presented in
Table~\ref{cel_ab}, we obtain ICF(He)$=1.63$.

For carbon, \ion{C}{ii} and \ion{C}{iii} ORLs have been detected. 
The ionic concentration in C$^{3+}$ should be negligible given the 
nearly absence of
He$^{2+}$ in \object{M\,2-24}. The C$^+$ ionic concentration needs to be 
accounted for.
Thus the carbon abundance is derived by 
\begin{eqnarray}
\frac{{\rm C}}{{\rm H}}&=&{\rm ICF}({\rm C})\times(\frac{{\rm C}^{2+}}{{\rm H}^+}+\frac{{\rm C}^{3+}}{{\rm H}^+}),\nonumber\\
\nonumber
\end{eqnarray}
where ICF is estimated as (Kingsburgh \& Barlow \cite{kingsburgh}),
\begin{eqnarray}
{\rm ICF(C)}=\frac{{\rm O}^{+}+{\rm O}^{2+}}{{\rm O}^{2+}}
            =1.04.\nonumber\\
\nonumber
\end{eqnarray}

For oxygen, the ionic concentration O$^{3+}$ again can be neglected. Thus the total elemental abundance of oxygen is
simply given by the sum of the ionic concentrations in O$^+$ and O$^{2+}$, 
both have been observed in the case of CEL analysis. For ORL analysis, however,
the O$^+$ ionic abundance cannot be determined. The total ORL abundance of 
oxygen thus is obtained by assuming that the O$^{2+}$/O ratio for the ORL
abundance is same as for the CEL abundance.

For nitrogen, neon, sulphur and argon, the total CEL abundances were derived
using the ICF formulae given by Kingsburgh \& Barlow (\cite{kingsburgh}). 
The total ORL abundances of nitrogen and neon are also estimated based on 
the assumption that the fractions of the unobserved ionic species for the ORL
abundances are the same as those in the case of the CEL abundances.
Among these elements, nitrogen has an extremely large
ICF of 23.3 because only singlet ionized stage is detected for this element.
Such a large ICF could lead to a large error in the final 
N/H elemental abundance ratio derived from CELs.

For magnesium, no ionic abundance is available from CELs and only
Mg$^{2+}$/H$^+$ abundance ratio has been determined from an ORL.
However, Mg$^{2+}$ is the dominant ionic stage of magnesium due to its large
ionization potential interval, ranging from 15 to 80\,eV. Mg$^{3+}$ must be
negligible in this low excitation nebula. Thus no ionization
correction is needed, and we have assumed that ${\rm Mg}/{\rm H}={\rm Mg}^{2+}/{\rm H}^+$.

For iron, we estimated the abundance by adding the ionic abundances of 
Fe$^{2+}$ and Fe$^{3+}$. Fe$^{4+}$ has been ignored based on the 
same reasoning for Mg$^{3+}$. The ionic concentration in Fe$^+$ was not
included as well because 
Fe$^0$ has a IP of only 7.9\,eV, thus Fe$^+$ exists mainly in the photodissociation 
regions outside the ionized zone (defined by H$^+$). 

The ORL analyses yield a He/H ratio of 0.107 and a N/O ratio of 0.25.
According to the criteria defined by Peimbert \&
Torres-Peimbert (\cite{peimbert}), \object{M\,2-24} is a Type-II
PN. Barlow et al. (\cite{barlow}) derived Mg/H abundances for ten
PNe and found they fall within a remarkably narrow range consistent
with the solar value. In \object{M\,2-24}, the Mg/H abundance is
higher than the solar value by a factor of 6.3, suggesting that not 
all PNe have
similar Mg/H abundances. The iron abundance in \object{M\,2-24}
is depleted by a factor of 26 with respect to the solar value. The
high depletion of iron in PNe is generally thought to be due to
the effective removal of gas phase iron by condensing into dust grains.

Given the insensitivity of the ORL abundances to temperature and density, 
and the very small (and similar) ICFs involved,
C/O, N/O, Ne/O and Mg/O ratios derived from
ORLs should be highly reliable. According to Table~\ref{abco}, the [C/O], 
[N/O], [Ne/O] and [Mg/O] 
ratios derived from ORLs are -0.08, 0.3, 0.62 and 0.47, respectively. The enhancement of
nitrogen could be attributed to the combined effects of the third
dredge-up with hot-bottom burning. It is however very interesting to note that two
$\alpha$-elements, neon and magnesium are overabundant with respect to oxygen
by large amounts. This is further discussed in the following section.

\section{Discussion}
 The optical spectrum of \object{M\,2-24} shows that it is a low-excitation PN. Based
on the excitation classification scheme proposed by Dopita \& 
Meatheringham (\cite{dopita90}), which makes use of the observed $I$([\ion{O}{iii}]$\lambda$5007)/$I( {\rm H}\beta)$
line ratio for low and intermediate excitation PNe (E.C. $\leq 5$), and the
$I({\rm He~II} \lambda4686)/I({\rm H}\beta)$ ratio for higher excitation class
PNe,
we find that \object{M\,2-24} has an E.C. of 1.7 from the
observed $I$([\ion{O}{iii}]$\lambda5007$)/$I({\rm H}\beta)$ ratio.

\begin{figure}
\centering
\epsfig{file=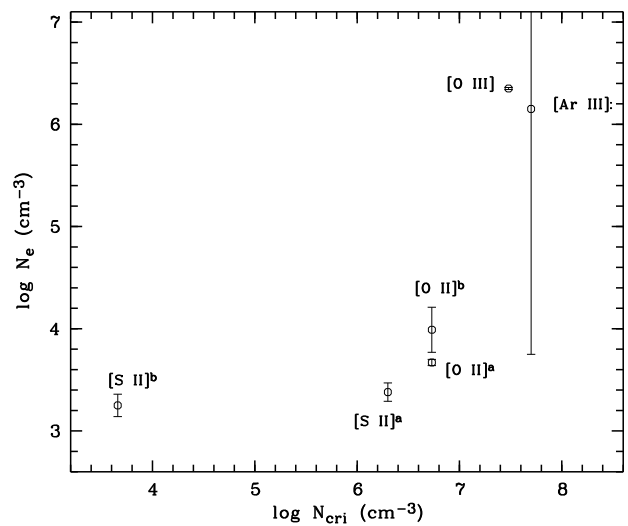, height=7.0cm,
bbllx=62, bblly=419, bburx=443, bbury=743, clip=, angle=0}
\caption{Electron density versus the critical densities
of these diagnostic line ratios.}
\label{crine}
\end{figure}

In Fig.~\ref{crine}, we show that in \object{M\,2-24}, the densities derived
from various diagnostic line ratios are positively correlated with their
critical densities. There might be a denser region in the nebular core, 
where all density-diagnostic CEL lines become invalid due to 
strong collisional de-excitation.
Such a high density nebular core might be caused by the strong mass exchange
between the central star and its companion or by compression of shock waves.
Similar features are also found in other PNe, such as 
\object{M\,2-9} (Allen \& Swings \cite{allen}), \object{Mz\,3} (Zhang \& Liu
\cite{zhangliu}), \object{He\,2-428} and \object{M\,1-91} (Rodr\'{i}guez
et al. \cite{rodriguez}). Zhang \& Liu (\cite{zhangliu}) showed that
[\ion{Fe}{iii}] forbidden lines are ideal diagnostics to probe the high
density regions because of their relative high critical densities
($\sim10^6$~cm$^{-3}$). 
Unfortunately, only a few [\ion{Fe}{iii}] lines are detected in the spectrum of 
\object{M\,2-24}.

\object{M\,2-24} shares some common characteristics with the bipolar
nebula \object{Mz\,3}. Both PNe are low excitation and have dense emission 
cores. For both PNe,
the extinctions yielded by the H$\alpha$/H$\beta$ ratio are significantly higher
than those derived from the higher order Balmer lines, suggesting that 
the Balmer lines from these dense central regions could be optically thick.
Thus, \object{M\,2-24} and \object{Mz\,3}
might share a similar evolutionary scenario that the two PNe may have been
produced by eruptions of binary
systems. On the other hand, there are some differences between the two objects.
\object{M\,2-24} has a prominently higher neon abundance than \object{Mz\,3}.
While the iron abundance of \object{M\,2-24} is relative lower than that of
\object{Mz\,3} (approx. one order of magnitude). This could imply that the
two objects have been formed in different chemical environments.

In our abundance analysis of the highly-ionized species, the Balmer jump temperature
$T_{\rm e}({\rm BJ})=16\,300$\,K was used. The ratio of the nebular continuum Balmer 
discontinuity to Balmer line is
practically independent of the density, thus $T_{\rm e}({\rm BJ})$ yields the average temperature
of the nebula. As a consequence, if there are large temperature variations in 
the nebula, our CEL abundances of high-ionization species will be problematic. Liu et al. 
(\cite{liu2001}) found the discrepancy between the ionic abundances derived from
ORL and CEL is positively correlated with the difference between the
electron temperatures derived from the [\ion{O}{iii}] forbidden line on the one
hand and from the nebular continuum Balmer discontinuity on the other, i.e.
$T_{\rm e}$([\ion{O}{iii}])$ - T_{\rm e}$(BJ). For \object{M\,2-24}, the 
O$^{2+}$/H$^+$ abundance ratio derived from ORLs is higher than that derived
from CELs by a factor of 17, suggesting $T_{\rm e}$([\ion{O}{iii}]) might be
higher than $T_{\rm e}$(BJ) by approximately about 5000\,K (c.f. the fitting 
given by Liu et al. \cite{liu2001}). Such a discrepancy of temperature is
partially attributed to the temperature fluctuation across the nebula, as suggested
by Peimbert (\cite{peimbert1971}), although the 
high-density regions can cause [\ion{O}{iii}] nebular
lines to be collisionally de-excited and lead to overestimated
$T_{\rm e}$([\ion{O}{iii}]), as pointed out by Viegas \& Clegg (\cite{viegas}).
Thus the adoption of $T_{\rm e}$(BJ) might lead to overestimating the CEL
abundances of \object{M\,2-24}. In order to clarify the effects of the
temperature fluctuation on CEL abundances in \object{M\,2-24}, better 
temperature-diagnostics for the high-density regions are needed.

In order to explain the disparity between the elemental abundances derived from
CEL and from ORL, 
Liu et al. (\cite{liubarlow}) presented a two-component nebular model,
with H-deficient material embedded in diffuse material with
`normal' abundance ($\sim$\,solar). The generally higher ORL abundances than the solar
suggest the H-deficient condensations might also exist in \object{M\,2-24}. 
ORL abundances of heavy elements have been enhanced by the H-deficient
inclusions. On the other hand, the presence of the high density core region
might lead to underestimated CEL abundances, as discussed above.
The lack of an ideal temperature-diagnostic for the 
high-density regions also causes some uncertainty in the derived CEL abundances.
Therefore, both the abundances derived here from CELs and from ORLs might be
uncertain for this particular PN. For the purpose of
accurate abundance determination, a proper photoionization
model, which includes a dense central emission core and possible existence
of H-deficient
condensations is needed, which is beyond the scope of the current paper.

It is noteworthy that this bulge PN has a large magnesium
abundance enhancement over solar. Barlow et al. (\cite{barlow}) have presented
that the observed enhancement in ORL abundance for second-row elements such
as carbon, nitrogen, oxygen and
neon, is absent for third-row elements such as magnesium and
silicon. Considering the depletion of magnesium in PNe, although 
expected to be small (Barlow et al. \cite{barlow}), the observed Mg/H 
ratio should represent a lower limit to the magnesium abundance of the
progenitor star of \object{M\,2-24}. Thus the high Mg/H ratio implies that the progenitor
star of this bulge PN may be extremely Mg-rich, suggesting it has formed in a 
very different environment with respect to the Sun. This is consistent with previous
studies which show that magnesium is generally enhanced in the Galactic bulge
(McWilliam \& Rich \cite{mcwilliam}). In addition, neon is 
enhanced by a large amount relative to the solar abundance. The general 
enhancement of $\alpha$-elements
in \object{M\,2-24} may suggest that Type {\sc II} supernova explosions
might have played a major role in the chemical evolution history of the 
Galactic bugle, given that $\alpha$-elements are mainly 
produced by massive stars which explode as Type {\sc II} supernovae.
To test this, however, the O, Ne and Mg abundances for a large sample of
Galactic bulge PNe need to be studied.
Alternatively, \object{M\,2-24} might have been heavily polluted by nucleosynthesis 
in its progenitor star. In other words, this object could be in fact a 
symbiotic nova. The presence of an
extremely dense core in \object{M\,2-24}, which might have been produced by,
e.g. mass transfer, seems
to be in favour of this argument. 
Further observation is needed to clarify this. 

\begin{acknowledgements}
  We are grateful to S.-G Luo and Y. Liu for their help with the preparation 
of this paper. We would also like to thank the anonymous referee for the
suggestions that improved this paper. This work was partially supported by 
Beijing Astrophysics Center (BAC).
\end{acknowledgements}

\end{document}

%% file: tab1.tex
 \begin{table*}
 \tabcolsep 10pt
 \centering
 \caption{\label{linelist} Observed                   relative line fluxes,                                             on a scale where H$\beta=100$.}
 \begin{tabular}{lccccccccc}
 \hline
 \hline
 \noalign{\smallskip}
 $\lambda_{\rm obs}$&$F(\lambda)$&                  $I(\lambda)$&Ion&$\lambda_{\rm lab}$                           &Mult&Lower Term&Upper Term&$g_1$&$g_2$\\
 (1) & (2) & (3) & (4) & (5) & (6)                     & (7) & (8)& (9)& (10) \\
 \noalign{\smallskip}
 \hline
 \noalign{\smallskip}
 3613.19&     0.266&     0.443&  He I    &  3613.64& V6     &   2s  1S  &  5p  1P*&  1&  3\\
 3634.05&     0.507&     0.837&  He I    &  3634.25& V28    &   2p  3P* &  8d  3D &  9& 15\\
 3673.79&     0.364&     0.593&  H 23    &  3673.74& H23    &   2p+ 2P* & 23d+ 2D &  8&  *\\
 3676.47&     0.502&     0.818&  H 22    &  3676.36& H22    &   2p+ 2P* & 22d+ 2D &  8&  *\\
 3679.46&     0.536&     0.873&  H 21    &  3679.36& H21    &   2p+ 2P* & 21d+ 2D &  8&  *\\
 3682.92&     0.603&     0.981&  H 20    &  3682.81& H20    &   2p+ 2P* & 20d+ 2D &  8&  *\\
 3686.80&     0.704&     1.143&  H 19    &  3686.83& H19    &   2p+ 2P* & 19d+ 2D &  8&  *\\
 3691.77&     0.872&     1.413&  H 18    &  3691.56& H18    &   2p+ 2P* & 18d+ 2D &  8&  *\\
 3697.29&     0.981&     1.590&  H 17    &  3697.15& H17    &   2p+ 2P* & 17d+ 2D &  8&  *\\
 3704.28&     1.900&     3.067&  H 16    &  3703.86& H16    &   2p+ 2P* & 16d+ 2D &  8&  *\\
       &         *&         *&  He I    &  3705.02& V25    &   2p  3P* &  7d  3D &  9& 15\\
 3707.79&     0.111&     0.179&  O III   &  3707.25& V14    &   3p  3P  &  3d  3D*&  3&  5\\
 3712.13&     1.520&     2.449&  H 15    &  3711.97& H15    &   2p+ 2P* & 15d+ 2D &  8&  *\\
       &         *&         *&  O III   &  3715.08& V14    &   3p  3P  &  3d  3D*&  5&  7\\
 3721.93&     2.680&     4.303&  H 14    &  3721.94& H14    &   2p+ 2P* & 14d+ 2D &  8&  *\\
       &         *&         *&  [S III] &  3721.63& F2     &   3p2 3P  &  3p2 1S &  3&  1\\
 3726.19&     6.090&     9.777&  [O II]  &  3726.03& F1     &   2p3 4S* &  2p3 2D*&  4&  4\\
 3728.48&     2.590&     4.150&  [O II]  &  3728.82& F1     &   2p3 4S* &  2p3 2D*&  4&  6\\
 3734.40&     1.930&     3.087&  H 13    &  3734.37& H13    &   2p+ 2P* & 13d+ 2D &  8&  *\\
 3750.19&     2.360&     3.761&  H 12    &  3750.15& H12    &   2p+ 2P* & 12d+ 2D &  8&  *\\
 3754.67&     0.296&     0.471&  O III   &  3754.69& V2     &   3s  3P* &  3p  3D &  3&  5\\
 3756.74&     0.170&     0.270&  O III   &  3757.24& V2     &   3s  3P* &  3p  3D &  1&  3\\
 3759.88&     0.160&     0.254&  O III   &  3759.87& V2     &   3s  3P* &  3p  3D &  5&  7\\
 3770.69&     2.890&     4.572&  H 11    &  3770.63& H11    &   2p+ 2P* & 11d+ 2D &  8&  *\\
 3774.17&     0.128&     0.202&  O III   &  3774.02& V2     &   3s  3P* &  3p  3D &  3&  3\\
 3777.64&     0.125&     0.197&  Ne II   &  3777.14& V1     &   3s  4P  &  3p  4P*&  2&  4\\
 3790.95&     0.071&     0.111&  O III   &  3791.27& V2     &   3s  3P* &  3p  3D &  5&  5\\
 3797.91&     3.720&     5.831&  H 10    &  3797.90& H10    &   2p+ 2P* & 10d+ 2D &  8&  *\\
 3806.57&     0.186&     0.290&  He I    &  3805.74& V63    &   2p  1P* & 11d  1D &  3&  5\\
 3814.25&     0.027&     0.042&  He II   &  3813.50&  4.19  &   4f+ 2F* & 19g+ 2G & 32&  *\\
 3819.69&     1.090&     1.696&  He I    &  3819.62& V22    &   2p  3P* &  6d  3D &  9& 15\\
 3835.37&     5.260&     8.139&  H 9     &  3835.39& H9     &   2p+ 2P* &  9d+ 2D &  8&  *\\
 3856.56&     0.158&     0.243&  O II    &  3856.13& V12    &   3p  4D* &  3d  4D &  4&  2\\
       &         *&         *&  Si II   &  3856.02& V1     &   3p2 2D  &  4p  2P*&  6&  4\\
 3859.06&     0.123&     0.189&  He II   &  3858.07&  4.17  &   4f+ 2F* & 17g+ 2G & 32&  *\\
 3862.63&     0.461&     0.707&  Si II   &  3862.60& V1     &   3p2 2D  &  4p  2P*&  4&  2\\
 3868.63&    95.900&   146.494&  [Ne III]&  3868.75& F1     &   2p4 3P  &  2p4 1D &  5&  5\\
 3882.09&     0.086&     0.131&  O II    &  3882.19& V12    &   3p  4D* &  3d  4D &  8&  8\\
       &         *&         *&  O II    &  3882.45& V11    &   3p  4D* &  3d  4P &  4&  4\\
       &         *&         *&  O II    &  3883.13& V12    &   3p  4D* &  3d  4D &  8&  6\\
 3888.93&     9.960&    15.103&  H 8     &  3889.05& H8     &   2p+ 2P* &  8d+ 2D &  8&  *\\
       &         *&         *&  He I    &  3888.65& V2     &   2s  3S  &  3p  3P*&  3&  9\\
 3926.64&     0.157&     0.235&  He I    &  3926.54& V58    &   2p  1P* &  8d  1D &  3&  5\\
 3967.40&    27.900&    41.153&  [Ne III]&  3967.46& F1     &   2p4 3P  &  2p4 1D &  3&  5\\
 3970.00&    10.800&    15.901&  H 7     &  3970.07& H7     &   2p+ 2P* &  7d+ 2D &  8& 98\\
 4009.49&     0.191&     0.277&  He I    &  4009.26& V55    &   2p  1P* &  7d  1D &  3&  5\\
 4026.27&     2.070&     2.981&  He I    &  4026.21& V18    &   2p  3P* &  5d  3D &  9& 15\\
       &         *&         *&  N II    &  4026.08& V39b   &   3d  3F* & 4f 2[5] &  7&  9\\
 4035.79&     0.062&     0.089&  N II    &  4035.08& V39a   &   3d  3F* & 4f 2[4] &  5&  7\\
 4041.47&     0.045&     0.065&  N II    &  4041.31& V39b   &   3d  3F* & 4f 2[5] &  9& 11\\
 4045.82&     0.087&     0.124&  [Fe III]&  4046.40&        &   3d6 5D  & 3d6 3G  &  7&  7\\
 4048.12&     0.038:&     0.054&  O II    &  4048.21& V50b   &   3d  4F  &  4f  F3*&  8&  8\\
 4062.83&     0.037:&     0.052&  O II    &  4062.94& V50a   &   3d  4F  &  4f  F4*& 10& 10\\
 4069.06&     1.340&     1.898&  [S II]  &  4068.60& F1     &   2p3 4S* &  2p3 2P*&  4&  4\\
 4072.02&     0.200&     0.283&  O II    &  4071.23& V48a   &   3d  4F  &  4f  G5*&  8& 10\\
       &         *&         *&  O II    &  4072.16& V10    &   3p  4D* &  3d  4F &  6&  8\\
 4076.30&     0.555&     0.783&  [S II]  &  4076.35& F1     &   2p3 4S* &  2p3 2P*&  2&  4\\
 4079.34&     0.050:&     0.071&  O II    &  4078.84& V10    &   3p  4D* &  3d  4F &  4&  4\\
 4088.89&     0.196&     0.275&  O II    &  4089.29& V48a   &   3d  4F  &  4f  G5*& 10& 12\\
 4097.42&     1.140&     1.597&  N III   &  4097.33& V1     &   3s  2S  &  3p  2P*&  2&  4\\
       &         *&         *&  O II    &  4097.25& V20    &   3p  4P* &  3d  4D &  2&  4\\
       &         *&         *&  O II    &  4097.26& V48b   &   3d  4F  &  4f  G4*&  8& 10\\
 \noalign{\smallskip}
 \hline
 \end{tabular}
 \end{table*}
 \setcounter{table}{1}
 \begin{table*}
 \tabcolsep 10pt
 \centering
 \caption{{\it --continued}}
 \begin{tabular}{lccccccccc}
 \hline
 \hline
 \noalign{\smallskip}
 $\lambda_{\rm obs}$&$F(\lambda)$&                      $I(\lambda)$&Ion&$\lambda_{\rm lab}$                           &Mult&Lower Term&Upper Term&$g_1$&$g_2$\\
 (1) & (2) & (3) & (4) & (5) & (6)                        & (7) & (8)& (9)& (10) \\
 \noalign{\smallskip}
 \hline
 \noalign{\smallskip}
       &         *&         *&  O II    &  4098.24& V46a   &   3d  4F  &  4f  D3*&  4&  6\\
 4101.81&    19.600&    27.407&  H 6     &  4101.74& H6     &   2p+ 2P* &  6d+ 2D &  8& 72\\
 4120.80&     0.466&     0.647&  O II    &  4119.22& V20    &   3p  4P* &  3d  4D &  6&  8\\
       &         *&         *&  O II    &  4120.28& V20    &   3p  4P* &  3d  4D &  6&  6\\
       &         *&         *&  O II    &  4120.54& V20    &   3p  4P* &  3d  4D &  6&  4\\
       &         *&         *&  He I    &  4120.84& V16    &   2p  3P* &  5s  3S &  9&  3\\
       &         *&         *&  O II    &  4121.46& V19    &   3p  4P* &  3d  4P &  2&  2\\
 4144.16&     0.699&     0.960&  He I    &  4143.76& V53    &   2p  1P* &  6d  1D &  3&  5\\
 4152.64&     0.167&     0.228&  O II    &  4153.30& V19    &   3p  4P* &  3d  4P &  4&  6\\
 4156.61&     0.075&     0.102&  O II    &  4156.53& V19    &   3p  4P* &  3d  4P &  6&  4\\
 4186.27&     0.120&     0.162&  C III   &  4186.90& V18    &   4f  1F* &  5g  1G &  7&  9\\
 4190.54&     0.087&     0.117&  O II    &  4189.79& V36    &   3p$'$ 2F* &  3d$'$ 2G &  8& 10\\
 4199.85&     0.213&     0.286&  He II   &  4199.83&  4.11  &   4f+ 2F* & 11g+ 2G & 32&  *\\
       &         *&         *&  N III   &  4200.10& V6     &   3s$'$ 2P* &  3p$'$ 2D &  4&  6\\
 4206.63&     0.141&     0.189&  [Fe IV] &  4206.60&        &   3d5 4G  &  3d5 2H & 10& 10\\
 4209.31&     0.120&     0.161&  [Fe IV] &  4208.90&        &   3d5 4G  &  3d5 2H &  8& 10\\
 4228.01&     0.040:&     0.053&  N II    &  4227.74& V33    &   3p  1D  &  4s  1P*&  5&  3\\
       &         *&         *&  [Fe V]  &  4227.20& F2     &   3d4 5D  & 3d4  3H &  9&  9\\
 4237.08&     0.046&     0.060&  N II    &  4236.91& V48a   &   3d  3D* & 4f 1[3] &  3&  5\\
       &         *&         *&  N II    &  4237.05& V48b   &   3d  3D* & 4f 1[4] &  5&  7\\
 4258.64&     0.054&     0.071&  [Fe III]&  4257.20&        &   3d6 3P  &  3d6 1F &  5&  7\\
 4267.46&     0.504&     0.657&  C II    &  4267.15& V6     &   3d  2D  &  4f  2F*& 10& 14\\
 4274.46&     0.046&     0.059&  O II    &  4273.10& V67a   &   3d  4D  &  4f  F4*&  6&  8\\
 4275.85&     0.029:&     0.038&  O II    &  4275.55& V67a   &   3d  4D  &  4f  F4*&  8& 10\\
       &         *&         *&  O II    &  4275.99& V67b   &   3d  4D  &  4f  F3*&  4&  6\\
       &         *&         *&  O II    &  4276.28& V67b   &   3d  4D  &  4f  F3*&  6&  6\\
       &         *&         *&  O II    &  4276.75& V67b   &   3d  4D  &  4f  F3*&  6&  8\\
 4277.56&     0.126&     0.164&  O II    &  4277.43& V67c   &   3d  4D  &  4f  F2*&  2&  4\\
       &         *&         *&  O II    &  4277.89& V67b   &   3d  4D  &  4f  F3*&  8&  8\\
 4280.61&     0.035&     0.046&  O II    &  4281.32& V53b   &   3d  4P  &  4f  D2*&  6&  6\\
 4283.10&     0.021&     0.028&  O II    &  4282.96& V67c   &   3d  4D  &  4f  F2*&  4&  6\\
 4284.67&     0.038&     0.050&  O II    &  4283.73& V67c   &   3d  4D  &  4f  F2*&  4&  4\\
 4286.85&     0.079&     0.102&  O II    &  4285.69& V78b   &   3d  2F  &  4f  F3*&  6&  8\\
 4289.54&     0.063&     0.081&  O II    &  4288.82& V53c   &   3d  4P  &  4f  D1*&  2&  4\\
       &         *&         *&  O II    &  4288.82& V53c   &   3d  4P  &  4f  D1*&  2&  2\\
 4292.26&     0.029&     0.038&  C II    &  4292.16&        &   4f  2F* & 10g  2G & 14& 18\\
       &         *&         *&  O II    &  4292.21& V78c   &   3d  2F  &  4f  F2*&  6&  6\\
 4317.45&     0.071&     0.091&  O II    &  4317.14& V2     &   3s  4P  &  3p  4P*&  2&  4\\
       &         *&         *&  O II    &  4317.70& V53a   &   3d  4P  &  4f  D3*&  4&  6\\
 4319.74&     0.042&     0.054&  O II    &  4319.63& V2     &   3s  4P  &  3p  4P*&  4&  6\\
 4333.12&     0.092&     0.117&  O II    &  4332.71& V65b   &   3d  4D  &  4f  G4*&  8& 10\\
 4340.56&    39.200&    49.532&  H 5     &  4340.47& H5     &   2p+ 2P* &  5d+ 2D &  8& 50\\
 4350.07&     0.220&     0.277&  O II    &  4349.43& V2     &   3s  4P  &  3p  4P*&  6&  6\\
 4352.94&     0.074&     0.093&  O II    &  4353.59& V76c   &   3d  2F  &  4f  G3*&  6&  8\\
 4363.21&    56.000&    69.982&  [O III] &  4363.21& F2     &   2p2 1D  &  2p2 1S &  5&  1\\
 4369.31&     0.052&     0.065&  Ne II   &  4369.86& V56    &   3d  4F  & 4f 0[3]*&  4&  6\\
 4378.91&     0.098&     0.121&  N III   &  4379.11& V18    &   4f  2F* &  5g  2G & 14& 18\\
       &         *&         *&  Ne II   &  4379.55& V60b   &   3d  3F  & 4f 1[4]*&  8& 10\\
 4388.24&     0.628&     0.776&  He I    &  4387.93& V51    &   2p  1P* &  5d  1D &  3&  5\\
 4391.41&     0.048&     0.059&  Ne II   &  4391.99& V55e   &   3d  4F  & 4f 2[5]*& 10& 12\\
       &         *&         *&  Ne II   &  4392.00& V55e   &   3d  4F  & 4f 2[5]*& 10& 10\\
 4413.12&     0.074&     0.091&  Ne II   &  4413.22& V65    &   3d  4P  & 4f 0[3]*&  6&  8\\
       &         *&         *&  Ne II   &  4413.11& V57c   &   3d  4F  & 4f 1[3]*&  4&  6\\
       &         *&         *&  Ne II   &  4413.11& V65    &   3d  4P  & 4f 0[3]*&  6&  6\\
 4415.60&     0.109&     0.133&  O II    &  4414.90& V5     &   3s  2P  &  3p  2D*&  4&  6\\
 4418.05&     0.180&     0.220&  O II    &  4416.97& V5     &   3s  2P  &  3p  2D*&  2&  4\\
 4428.13&     0.071&     0.087&  Ne II   &  4428.64& V60c   &   3d  2F  & 4f 1[3]*&  6&  8\\
       &         *&         *&  Ne II   &  4428.52& V61b   &   3d  2D  & 4f 2[3]*&  6&  8\\
 4432.97&     0.028&     0.034&  N II    &  4432.74& V55a   &   3d  3P* & 4f 2[3] &  5&  7\\
       &         *&         *&  N II    &  4433.48& V55b   &   3d  3P* & 4f 2[2] &  1&  3\\
 4438.15&     0.109:&     0.132&  He I    &  4437.55& V50    &   2p  1P* &  5s  1S &  3&  1\\
 4471.69&     5.780&     6.885&  He I    &  4471.50& V14    &   2p  3P* &  4d  3D &  9& 15\\
 \noalign{\smallskip}
 \hline
 \end{tabular}
 \end{table*}
 \setcounter{table}{1}
 \begin{table*}
 \tabcolsep 10pt
 \centering
 \caption{{\it --continued}}
 \begin{tabular}{lccccccccc}
 \hline
 \hline
 \noalign{\smallskip}
 $\lambda_{\rm obs}$&$F(\lambda)$&                      $I(\lambda)$&Ion&$\lambda_{\rm lab}$                           &Mult&Lower Term&Upper Term&$g_1$&$g_2$\\
 (1) & (2) & (3) & (4) & (5) & (6)                        & (7) & (8)& (9)& (10) \\
 \noalign{\smallskip}
 \hline
 \noalign{\smallskip}
 4481.77&     0.291&     0.345&  Mg II   &  4481.21& V4     &   3d  2D  &  4f  2F*& 10& 14\\
 4491.82&     0.070&     0.083&  C II    &  4491.07&        &   4f  2F* &  9g  2G & 14& 18\\
       &         *&         *&  O II    &  4491.23& V86a   &   3d  2P  &  4f  D3*&  4&  6\\
 4511.03&     0.156&     0.183&  N III   &  4510.91& V3     &   3s$'$ 4P* &  3p$'$ 4D &  4&  6\\
       &         *&         *&  N III   &  4510.91& V3     &   3s$'$ 4P* &  3p$'$ 4D &  2&  4\\
 4514.75&     0.041&     0.048&  N III   &  4514.86& V3     &   3s$'$ 4P* &  3p$'$ 4D &  6&  8\\
 4517.53&     0.077&     0.090&  N III   &  4518.15& V3     &   3s$'$ 4P* &  3p$'$ 4D &  2&  2\\
 4523.33&     0.080&     0.093&  N III   &  4523.58& V3     &   3s$'$ 4P* &  3p$'$ 4D &  4&  4\\
 4530.00&     0.363&     0.421&  N II    &  4530.41& V58b   &   3d  1F* & 4f 2[5] &  7&  9\\
       &         *&         *&  N III   &  4530.86& V3     &   3s$'$ 4P* &  3p$'$ 4D &  4&  2\\
 4592.20&     0.273&     0.308&  O II    &  4590.97& V15    &   3s$'$ 2D  &  3p$'$ 2F*&  6&  8\\
 4596.65&     0.148&     0.167&  O II    &  4596.18& V15    &   3s$'$ 2D  &  3p$'$ 2F*&  4&  6\\
 4607.29&     0.072&     0.081&  N II    &  4607.16& V5     &   3s  3P* &  3p  3P &  1&  3\\
 4609.88&     0.113&     0.127&  O II    &  4609.44& V92a   &   3d  2D  &  4f  F4*&  6&  8\\
       &         *&         *&  O II    &  4610.20& V92c   &   3d  2D  &  4f  F2*&  4&  6\\
 4634.53&     0.871&     0.966&  N III   &  4634.14& V2     &   3p  2P* &  3d  2D &  2&  4\\
 4641.25&     2.020&     2.231&  N III   &  4640.64& V2     &   3p  2P* &  3d  2D &  4&  6\\
       &         *&         *&  O II    &  4641.81& V1     &   3s  4P  &  3p  4D*&  4&  6\\
       &         *&         *&  N III   &  4641.84& V2     &   3p  2P* &  3d  2D &  4&  4\\
 4649.71&     1.300&     1.431&  O II    &  4649.13& V1     &   3s  4P  &  3p  4D*&  6&  8\\
       &         *&         *&  C III   &  4650.25& V1     &   3s  3S  &  3p  3P*&  3&  3\\
       &         *&         *&  O II    &  4650.84& V1     &   3s  4P  &  3p  4D*&  2&  2\\
       &         *&         *&  C III   &  4651.47& V1     &   3s  3S  &  3p  3P*&  3&  1\\
 4659.10&     0.372&     0.408&  [Fe III]&  4658.10& F3     &   3d6 5D  &  3d6 3F2&  9&  9\\
       &         *&         *&  C IV    &  4657.15&        &   5f  2F* &  6g  2G & 14& 18\\
 4662.24&     0.094&     0.102&  O II    &  4661.63& V1     &   3s  4P  &  3p  4D*&  4&  4\\
 4676.54&     0.161&     0.175&  O II    &  4676.24& V1     &   3s  4P  &  3p  4D*&  6&  6\\
 4677.79&     0.050&     0.054&  N II    &  4678.14& V61b   &   3d  1P* & 4f 2[2] &  3&  5\\
 4686.37&     0.516&     0.559&  He II   &  4685.68&  3.4   &   3d+ 2D  &  4f+ 2F*& 18& 32\\
 4701.65&     0.099&     0.107&  [Fe III]&  4701.62& F3     &   3d6 5D  &  3d6 3F2&  7&  7\\
 4713.46&     1.240&     1.327&  [Ar IV] &  4711.37& F1     &   3p3 4S* &  3p3 2D*&  4&  6\\
       &         *&         *&  He I    &  4713.17& V12    &   2p  3P* &  4s  3S &  9&  3\\
 4733.90&     0.046&     0.049&  [Fe III]&  4733.87&        &   3d6 5D  &  3d6 3F &  5&  5\\
 4740.55&     1.380&     1.458&  [Ar IV] &  4740.17& F1     &   3p3 4S* &  3p3 2D*&  4&  4\\
 4860.96&   100.000&   100.000&  H 4     &  4861.33& H4     &   2p+ 2P* &  4d+ 2D &  8& 32\\
 4881.98&     0.216&     0.214&  [Fe III]&  4881.11& F2     &   3d6 5D  &  3d6 3H &  9&  9\\
 4904.52&     0.810&     0.800&  O II    &  4906.83& V28    &   3p  4S* &  3d  4P &  4&  4\\
       &         *&         *&  [Fe II] &  4905.34&        &   3d7 4F  &  4s  4F &  8&  8\\
       &         *&         *&  [Fe IV] &  4906.60&        &   3d5 4G  &  3d5 4F & 12& 10\\
 4921.58&     1.650&     1.605&  He I    &  4921.93& V48    &   2p  1P* &  4d  1D &  3&  5\\
 4958.55&   128.000&   122.464&  [O III] &  4958.91& F1     &   2p2 3P  &  2p2 1D &  3&  5\\
 5006.51&   394.000&   368.720&  [O III] &  5006.84& F1     &   2p2 3P  &  2p2 1D &  5&  5\\
 5032.60&     0.519&     0.480&  [Fe IV] &  5032.40&        &   3d5 4G  &  3d5 2F &  6&  6\\
 5041.48&     1.780&     1.641&  Si II   &  5041.03& V5     &   4p  2P* &  4d  2D &  2&  4\\
 5056.10&     0.657&     0.601&  Si II   &  5055.98& V5     &   4p  2P* &  4d  2D &  4&  6\\
       &         *&         *&  Si II   &  5056.31& V5     &   4p  2P* &  4d  2D &  4&  4\\
 5191.28&     0.226:&     0.195&  [Ar III]&  5191.82& F3     &   2p4 1D  &  2p4 1S &  5&  1\\
 5199.63&     0.580&     0.498&  [N I]   &  5199.84& F1     &   2p3 4S* &  2p3 2D*&  4&  4\\
       &         *&         *&  [N I]   &  5200.26& F1     &   2p3 4S* &  2p3 2D*&  4&  6\\
 5341.69&     0.095&     0.077&  C II    &  5342.38&        &   4f  2F* &  7g  2G & 14& 18\\
 5679.07&     0.304&     0.217&  N II    &  5679.56& V3     &   3s  3P* &  3p  3D &  5&  7\\
 5754.46&     1.820&     1.271&  [N II]  &  5754.60& F3     &   2p2 1D  &  2p2 1S &  5&  1\\
 5798.93&     0.379&     0.261&  C IV    &  5801.51& V1     &   3s  2S  &  3p  2P*&  2&  4\\
 5810.84&     0.296&     0.203&  C IV    &  5812.14& V1     &   3s  2S  &  3p  2P*&  2&  2\\
 5875.63&    30.700&    20.660&  He I    &  5875.66& V11    &   2p  3P* &  3d  3D &  9& 15\\
 5939.31&     0.215&     0.142&  N II    &  5940.24& V28    &   3p  3P  &  3d  3D*&  3&  3\\
       &         *&         *&  N II    &  5941.65& V28    &   3p  3P  &  3d  3D*&  5&  7\\
 6101.74&     0.859&     0.541&  [K IV]  &  6101.83& F1     &   3p4 3P  &  3d4 1D &  5&  5\\
 6301.51&     3.860&     2.296&  [O I]   &  6300.34& F1     &   2p4 3P  &  2p4 1D &  5&  5\\
 6312.30&     4.320&     2.565&  [S III] &  6312.10& F3     &   2p2 1D  &  2p2 1S &  5&  1\\
 6347.27&     0.470&     0.276&  Si II   &  6347.10& V2     &   4s  2S  &  4p  2P*&  2&  4\\
 6364.79&     1.250&     0.731&  [O I]   &  6363.78& F1     &   2p4 3P  &  2p4 1D &  3&  5\\
 \noalign{\smallskip}
 \hline
 \end{tabular}
 \end{table*}
 \setcounter{table}{1}
 \begin{table*}
 \tabcolsep 10pt
 \centering
 \caption{{\it --continued}}
 \begin{tabular}{lccccccccc}
 \hline
 \hline
 \noalign{\smallskip}
 $\lambda_{\rm obs}$&$F(\lambda)$&                      $I(\lambda)$&Ion&$\lambda_{\rm lab}$                           &Mult&Lower Term&Upper Term&$g_1$&$g_2$\\
 (1) & (2) & (3) & (4) & (5) & (6)                        & (7) & (8)& (9)& (10) \\
 \noalign{\smallskip}
 \hline
 \noalign{\smallskip}
 6371.94&     0.631&     0.368&  Si II   &  6371.38& V2     &   4s  2S  &  4p  2P*&  2&  2\\
 6549.53&    35.700&    19.873&  [N II]  &  6548.10& F1     &   2p2 3P  &  2p2 1D &  3&  5\\
 6562.78&   616.000&   341.649&  H 3     &  6562.77& H3     &   2p+ 2P* &  3d+ 2D &  8& 18\\
 6584.85&   109.000&    60.121&  [N II]  &  6583.50& F1     &   2p2 3P  &  2p2 1D &  5&  5\\
 6678.55&     8.300&     4.470&  He I    &  6678.16& V46    &   2p  1P* &  3d  1D &  3&  5\\
 6718.03&     9.840&     5.241&  [S II]  &  6716.44& F2     &   2p3 4S* &  2p3 2D*&  4&  6\\
 6732.54&    13.000&     6.898&  [S II]  &  6730.82& F2     &   2p3 4S* &  2p3 2D*&  4&  4\\
 7064.86&    18.500&     9.069&  He I    &  7065.25& V10    &   2p  3P* &  3s  3S &  9&  3\\
 7136.23&    13.400&     6.461&  [Ar III]&  7135.80& F1     &   3p4 3P  &  3p4 1D &  5&  5\\
 7159.25&     0.725&     0.348&  He I    &  7160.56&        &   3s  3S  & 10p  3P*&  3&  9\\
 7169.69&     1.410&     0.674&  [Ar IV] &  7170.62& F2     &   3p3 2D* &  3p3 2P*&  4&  4\\
 7236.13&     1.370&     0.646&  C II    &  7236.42& V3     &   3p  2P* &  3d  2D &  4&  6\\
       &         *&         *&  C II    &  7237.17& V3     &   3p  2P* &  3d  2D &  4&  4\\
       &         *&         *&  [Ar IV] &  7237.26& F2     &   3p3 2D* &  3p3 2P*&  6&  4\\
 7262.24&     0.991&     0.464&  [Ar IV] &  7262.76& F2     &   3p3 2D* &  3p3 2P*&  4&  2\\
 7280.34&     1.430&     0.667&  He I    &  7281.35& V45    &   2p  1P* &  3s  1S &  3&  1\\
 7296.07&     0.454:&     0.211&  He I    &  7298.04&        &   3s  3S  &  9p  3P*&  3&  9\\
 7320.07&     1.800&     0.833&  [O II]  &  7318.92& F2     &   2p3 2D* &  2p3 2P*&  6&  2\\
       &         *&         *&  [O II]  &  7319.99& F2     &   2p3 2D* &  2p3 2P*&  6&  4\\
 7330.40&     2.240&     1.033&  [O II]  &  7329.67& F2     &   2p3 2D* &  2p3 2P*&  4&  2\\
       &         *&         *&  [O II]  &  7330.73& F2     &   2p3 2D* &  2p3 2P*&  4&  4\\
 \noalign{\smallskip}
 \hline
 \end{tabular}
 \end{table*}